\documentstyle[12pt]{article}
\newcommand{\BEQ}{\begin{equation}}    % Gleichungen Anfang ..
\newcommand{\BEA}{\begin{eqnarray}}
\newcommand{\EEQ}{\end{equation}}      % .. und Ende
\newcommand{\EEA}{\end{eqnarray}}
\newcommand{\eps}{\epsilon}                      % epsilon
\newcommand{\lmb}{\lambda}                       % lambda
\newcommand{\sig}{\sigma}                        % sigma
\newcommand{\rar}{\rightarrow}                   % Pfeil nach rechts
\newcommand{\im}{\imath}                         % i ohne Punkt (Math.)
\newcommand{\jm}{\jmath}                         % j ohne Punkt (Math.)
\newcommand{\zeile}[1]{\vskip #1 \baselineskip}  % N Zeilen ueberschlagen
\newcommand{\vekz}[2]
     {\mbox{${\begin{array}{c} #1  \\ #2 \end{array}}$}}
\newcommand{\build}[3]{\mathrel{\mathop{\kern 0pt#1}\limits_{#2}^{#3} }}
\newcommand{\appsection}[2]{\setcounter{equation}{0} \section*{Appendix #1. #2}
\renewcommand{\theequation}{#1.\arabic{equation}}
              \renewcommand{\thesection}{#1} }
\catcode`\@=11
\def\numberbysection{\@addtoreset{equation}{section}
        \def\theequation{\thesection.\arabic{equation}}}
\numberbysection

\begin{document}
\baselineskip 0.3in
%
%Titelseite
%
\begin{titlepage}
\begin{flushright}
 OUTP-93-33S \\
 UGVA-DPT 1993/09-833
\end{flushright}
%\vskip 1cm
\begin{center}
%~\hfill Oktober 1993\\
\vskip 0.5in
{\Large \bf Schr\"odinger Invariance and Strongly Anisotropic Critical
Systems}
\vskip 0.5in
Malte Henkel \\[.3in]
{\em Department of Physics, Theoretical Physics, \\
University of Oxford, 1 Keble Road,
Oxford OX1 3NP, UK}\footnote{present adress}
 \\
and \\
{\em D\'epartement de Physique Th\'{e}orique,
     Universit\'e de Gen\`eve \\
     24  quai Ernest Ansermet,
     CH - 1211 Gen\`eve 4, Switzerland}
\end{center}
\zeile{2}
%%
% Abstract
%
\begin{abstract}
%{\bf Abstract} \zeile{1}
The extension of strongly anisotropic or dynamical scaling to
local scale invariance is investigated.
For the special case of an anisotropy or dynamical exponent $\theta=z=2$,
the group of local scale transformation considered is the Schr\"odinger group,
which can be obtained as the non-relativistic limit of the conformal group.
The requirement of Schr\"odinger invariance determines the two-point
function in the bulk and reduces the three-point function to a scaling
form of a single variable. Scaling forms are also derived
for the two-point function
close to a free surface which can be either space-like or time-like.
These results are reproduced in several exactly solvable statistical systems,
namely the kinetic Ising model with Glauber dynamics, lattice diffusion,
Lifshitz points in the spherical model and critical dynamics of the
spherical model with a non-conserved order parameter.

For generic values of $\theta$, evidence from higher order Lifshitz points
in the spherical model and from directed percolation suggests a simple
scaling form of the two-point function.
\end{abstract}

\end{titlepage}

\newpage
%
%Text der Arbeit
%
\section{Introduction}

Scale invariance is a central notion in present
theories of critical behaviour. In the context of two-dimensional,
static and isotropic critical behaviour, these
ideas have become spectactularly succesful in the
context of conformal invariance
\cite{Bela84}. The main physical idea behind this is the extension of
the covariance of correlation functions under length rescaling by a
constant factor $\lambda$ to general, space-dependent rescalings
$\lambda(\vec{r}\,)$. A coordinate transformation
$\lmb(\vec{r}\,)$ is conformal
if the angles are kept unchanged. In two dimensions, the conformal
group is infinite-dimensional.
This has lead, for example, to the exact calculation of critical
exponents and correlation functions and yields a
handle for classifying two-dimensional
universality classes (for reviews, see e.g. \cite{Card90}).

Much less is known about non-isotropic scaling. Consider a (connected)
correlation function $C(\vec{r};t)$
depending on ``space'' coordinates $\vec{r}$ and a
``time'' coordinate $t$ which satisfies the scaling relation
\BEQ \label{eq:AIS}
C( \lmb \vec{r}; \lmb^\theta t) = \lmb^{-2 x} C(\vec{r}; t)
\EEQ
where $x$ is a scaling dimension and $\theta$ is referred
to as {\em anisotropy exponent}.
Systems which satisfy eq.~(\ref{eq:AIS}) with $\theta \neq 1$
are by definition {\em strongly anisotropic
critical systems}. In fact, dynamical scaling of this kind
apears quite commonly in time-delayed averages close to an
equilibrium phase transition, where the anisotropy exponent is referred
to as dynamical exponent $z=\theta$, see \cite{Ferr67,Hohe77},
or in domain growth problems of systems quenched below the
equilibrium critical point, see \cite{Bind74,Jans89} and
\cite{Bray93} for a recent review. Alternatively, strongly anisotropic
scaling may arise in statics, examples are provided by directed
percolation \cite{Kinz83} or by magnetic systems
at a Lifshitz point \cite{Horn75}, where the anisotropy exponent
$\theta=\nu_{\|}/\nu_{\perp}$ is related to the
critical exponents $\nu_{\|,\perp}$ of the correlation lengths
$\xi_{\|,\perp}$ parallel and
perpendicular to the preferred direction.
Eq.~(\ref{eq:AIS}) can be recast in the form
\BEQ \label{eq:ScalForm}
C(\vec{r};t) = t^{-2x/\theta} \Phi\left( u \right)
\EEQ
which defines the scaling function $\Phi(u)$ and
\BEQ
u = \frac{r^{\theta}}{t}
\EEQ
is the scaling variable and henceforth we shall be always taking the scaling
limit $r\rar\infty , t\rar\infty$ where $u$ is kept fixed.

We ask the following question: what can be said
about the scaling function $\Phi(u)$?
Is it sensible to look beyond global
scaling with $\lmb$ constant to a space-time
dependent rescaling factor $\lmb(\vec{r},t)$?

This question had been adressed by Cardy
\cite{Card85}. Assuming dynamical scaling
for the dynamic response function, he takes as
extended set of local scale transformations
$\lmb(\vec{r}\,)$ the space-dependent
scaling $\vec{r}\rar\lmb(\vec{r}\,) \vec{r},
t \rar \lmb(\vec{r}\,)^{\theta} t$ where the $\lmb(\vec{r}\,)$ are
two-dimensional conformal transformations. This means that only
systems at a static critical point are considered. The assumed covariance
of the response function is used to map the problem from the
two-dimensional plane (in the space coordinates) to the strip geometry,
with a non-uniform rate. Next, since close to a static critical point the
static correlation length of the system in the strip is of the same order
of magnitude as its width, it is claimed that
{\it ``\ldots on much larger distances it is
permissible to use mean field theory to
calculate the dynamic correlation function in the strip''}
\cite{Card85}. For a system
with a non-conserved order parameter the response function then turns
out to be \cite{Card85}
\BEQ \label{eq:CRF}
G(r,t) \sim t^{-2x/\theta -1}\exp\left(-\frac{r^{\theta}}{t}\right)
\EEQ
where some non-universal constants have been suppressed.
The case of a conserved order
parameter was also treated. While this result is
appealingly simple, the assumptions
made in deriving it may appear to be quite strong,\footnote{The restriction
to $2D$ space is not really required and could be removed \cite{Card93}}
in particular the use of
mean-field (van Hove) theory. Also, one might wish to reconsider
the assumption that $\lmb(\vec{r}\,)$ were time-independent.
In fact, we shall study the scaling of the two-point function in
$(1+1)D$ directed percolation and higher
order Lifshitz points in the spherical model, both with $\theta\neq 2$,
and find that the form of the scaling function of these models does
not agree with eq.~(\ref{eq:CRF}).

Here we propose another group of space-time dependent local
scalings $\lmb(\vec{r},t)$. For definiteness,
we shall only consider the case $\theta=2$, but we do not have to make the
restriction to $d=2$ space dimensions. Although the van Hove theory for
a non-conserved order parameter has also $\theta=2$ (see \cite{Ma76}),
we do not make any of the approximations involved in that theory.
Rather, it is our aim to find the scaling functions
merely from their transformation properties under local scale transformations.
By taking $\theta=2$, we mean to perform the simplest case study of local,
albeit not conformal, scaling transformations.
The group of the local scaling transformations is the Schr\"odinger group,
which shall be defined in the next section. The approach chosen
has the advantage of being close in spirit to the
earliest investigations of conformal
invariance in critical phenomena \cite{Poly70}. On a more formal level, the
comparison of conformal with Schr\"odinger invariance provides some insight
into the characteristic properties of both.

Our results are as follows:
\begin{enumerate}
\item If the domain of both time and space coordinates is
infinite in extent and the scaling fields transform covariantly under the
Schr\"odinger group, the
two-point function is completely determined, while the three-point function is
reduced to a scaling form of one variable, see
eqs.~(\ref{ZweiPunkt},\ref{DreiPunkt}). For example, this applies to the
calculation of time-delayed correlations of systems at
equilibrium and at a static critical point, or else to
lattice diffusion problems.
\item If the space geometry is semi-infinite, a scaling form for the two-point
function will be derived, see eq.~(\ref{SOber}). This may be relevant to
critical dynamics close to a surface, see \cite{Droz91} for an example.
\item For a system being in a predefined initial state, it can be shown that
critical relaxation towards equilibrium displays dynamical scaling already
at intermediate times \cite{Jans89}, much later than microscopic times but
also well before the late time regime usually considered.
We derive a scaling form for
the two-point function, see eq.~(\ref{TOber}).
\item These results can be reproduced from a variety of exactly solvable
models.
\item The Schr\"odinger group can be extended to an infinite-dimensional group,
whose Lie algebra contains a Virasoro subalgebra. It can be
shown that for systems
with local interactions Schr\"odinger invariance follows from the requirements
of translation invariance in both space and time, rotation invariance in space,
scale invariance and Galilei invariance. If no anomalies occur, this even holds
for the whole infinite-dimensional group.
\end{enumerate}

The work described in this paper uses background from both conformal
field theory and time-dependent statistical mechanics. To make the
paper accessible to readers with knowledge in one, but not both of these
fields, we repeat in section~2 the definition of the Schr\"odinger group
and recall a few well-known facts about Galilei-invariant theories and
dynamical scaling. Section~3 describes
the derivation of the two- and three-point functions for either
infinite or semi-infinite geometries. The Schr\"odinger Ward identity is
considered as well, and the non-existence of non-conventional central
extensions
of the Schr\"odinger Lie algebra is shown (Appendix~B).
The discussions of this section
follow closely the known derivation of correlation functions from
conformal invariance. In section~4, we test and confirm the predictions from
Schr\"odinger invariance by calculating two- and three-point functions in
several exactly solvable and strongly anisotropic critical models.
In section~5,
we examine the scaling of the two-point function for some systems with
an anisotropy exponent $\theta \neq 2$. Exact and numerical results indicate
a disagreement with that result eq.~(\ref{eq:CRF}) obtained from $2D$ conformal
invariance and suggest an alternative simple scaling form. Section~6 gives
our conclusions.

\section{Basic concepts and terminology}

We begin by recalling some well-known facts about the
Schr\"odinger group, Galilean
invariance in field theories and dynamical scaling.

\subsection{The Schr\"odinger group}

The Schr\"odinger group is defined \cite{Nied72,Hage72} by
the following set of transformations
\BEQ \label{eq:SCHgr}
\vec{r}\rar \vec{r}\,' = \frac{{\cal R}\vec{r}+
\vec{v}\, t+\vec{a}}{\gamma t +\delta}
\;\; , \;\; t\rar t' =
\frac{\alpha t + \beta}{\gamma t+\delta} \;\; ; \;\;
\alpha \delta - \beta \gamma = 1
\EEQ
where $\alpha,\beta,\gamma,\delta,\vec{v},\vec{a}$ are real parameters and
$\cal R$ is a rotation matrix in $d$ space dimensions.
It is apparent that the Schr\"odinger
group can be obtained from the Galilei group
by extending the time translations to the full
M\"obius group $Sl(2,R)$ of fractional real linear transformations
in time as given in eq.~(\ref{eq:SCHgr}). A faithful
matrix representation is given by
\BEQ
{\cal L}_g = \left( \begin{array}{ccc}
{\cal R} & \vec{v} & \vec{a} \\
   0     & \alpha  & \beta   \\
   0     & \gamma  & \delta \end{array} \right)
\;\; , \;\; {\cal L}_{g} {\cal L}_{g'} = {\cal L}_{g g'}
\EEQ

Niederer \cite{Nied72} showed that this group
is the maximal kinematical group which transforms solutions of the free
Schr\"odinger equation
\BEQ \label{eq:FSCH}
\left( i \frac{\partial}{\partial t} +
\frac{1}{2m}\frac{\partial^2}{\partial r^2}
\right) \psi = 0
\EEQ
into other solutions of (\ref{eq:FSCH}),
viz. $(\vec{r},t) \mapsto g(\vec{r},t)$,
$\psi \rar T_g \psi$
\BEQ \label{eq:wave}
(T_g \psi)(\vec{r},t) = f_g \left( g^{-1}(\vec{r},t)\right)
\psi \left( g^{-1}(\vec{r},t)\right)
\EEQ
where the companion function $f_g$ is \cite{Nied72}
\BEQ
f_g\left( \vec{r},t\right) = (\gamma t +\delta)^{-d/2}
\exp \left[ -\frac{im}{2}
\frac{\gamma {\vec{r}\,}^2 + 2
{\cal R}\vec{r}\cdot\left( \gamma \vec{a}-\delta \vec{v}\,\right)
+\gamma {\vec{a}\,}^2 -\delta t {\vec{v}\,}^2 +
2\gamma \vec{a}\vec{v}\,}{\gamma t +\delta}\right]
\EEQ

Independently, it was shown by Hagen \cite{Hage72} that non-relativistic free
field theory is Schr\"odinger invariant, treating both scalar and
spin-$\frac{1}{2}$ fields. It was also shown that the
operators which appear in the conservations laws associated with the
space-time symmetries can be reformulated to allow the statement of
Schr\"odinger invariance entirely in terms of those operator
densities \cite{Hage72}. The Schr\"odinger group
in $d$ space dimensions can be
obtained by a group contraction (where the speed of
light $c\rar \infty$) from the conformal group in $d+1$ dimensions
\cite{Baru73} provided the mass is conviently
rescaled as well. Its projective representations as required by (\ref{eq:wave})
have been studied in detail \cite{Perr77}.

There are many more equations, whose kinematical group is isomorphic to the
Schr\"odinger group. It can be shown that the most general potential which can
be added in (\ref{eq:FSCH}) such that the kinematical group is still isomorphic
to (\ref{eq:SCHgr}) is of the form, up to orthogonal transformations
$V(\vec{r}\,) = v^{(2)} \sum_{i} r_{i}^2 +
\sum_i v_{i}^{(1)} r_i + v^{(0)}$
\cite{Nied74}. Further examples are provided
by a non-linear Schr\"odinger equation or, but with a more general
transformation law than (\ref{eq:wave}), by the Navier-Stokes equation
with homogenous pressure or by Burger's
equation \cite{Nied78}. Higher order symmetry operators
of eq.~(\ref{eq:FSCH}) are examined in \cite{Beck91}.
The Schr\"odinger group also appears as a dynamical symmetry group
for the Dirac monopole or magnetic vortices \cite{Jack80}, or in the
non-relativistic $N-$ body problem with inverse-square
interactions \cite{Duva91}.
Similarly, one may treat the
diffusion equation by writing $m^{-1} = 2 i D$, where $D$ is
the diffusion constant. We shall do
so for most of this paper. In any case, we shall only
consider here the realization of
the Schr\"odinger group provided by the free
Schr\"odinger equation (\ref{eq:FSCH}).

For simplicity, we restrict attention here to
fields which are scalar under space
rotations (so that it is sufficient to take
${\cal R}=1$) and shall mostly also take
just one space dimension $d=1$. This is not a
serious restriction and generalizations
are straightforward.

The set ${\cal S}_{\rm fin} =
\{ X_{-1},X_{0},X_{1},Y_{-1/2},Y_{1/2},M_0 \}$ spans
the Lie algebra of the Schr\"o\-din\-ger group eq.~(\ref{eq:SCHgr}).
The generators read (we take $d=1$)
\BEA
X_n &=& -t^{n+1}\partial_t -\frac{n+1}{2} t^n r\partial_r -\frac{n(n+1)}{4}
{\cal M} t^{n-1} r^2 \nonumber \\
Y_m &=& -t^{m+1/2}\partial_r -\left( m+\frac{1}{2}\right) {\cal M} t^{m-1/2} r
\nonumber \\
M_n &=& -t^n {\cal M} \label{eq:SGen}
\EEA
where the terms $\sim {\cal M}$ come from the companion function.
When ${\cal M}= i m$ is purely imaginary, this corresponds to the Schr\"odinger
equation where $m$ is the mass, while for $\cal M$ real, this is the form
corresponding to the diffusion equation.
The commutation relations are
\BEA \label{eq:SCHComm}
\left[ X_n , X_m \right] &=& (n-m) X_{n+m} \nonumber \\
\left[ X_n , Y_m \right] &=& \left( \frac{n}{2} - m\right) Y_{n+m} \nonumber \\
\left[ X_n , M_m \right] &=& -m M_{n+m} \nonumber \\
\left[ Y_n , Y_m \right] &=& (n-m) M_{n+m} \nonumber \\
\left[ Y_n , M_m \right] &=& \left[ M_n , M_m \right] = 0
\EEA
(In more than one space dimensions, there are several sets of generators
$Y_{m}^{(i)}, i=1,\ldots,d$, but only one set of $X_n$, where $r\partial_r$
is replaced by $\vec{r}\partial_{\vec{r}}$ etc.) The commutation
relations (\ref{eq:SCHComm}) remain valid when the infinite set of generators
${\cal S}=\{ X_n, Y_m, M_n \}$ where $n$ is an integer and
$m$ is a half-integer, is considered \cite{Henk92}. The Lie algebra
can be decomposed ${\cal S}={\cal S}_X \oplus_s {\cal S}_Y$ where
${\cal S}_X = \{ X_n \}$ and ${\cal S}_Y = \{ Y_m, M_n\}$. As we shall see
later, these two subalgebras arise in quite distinct physical situations.

\subsection{Galilei-invariant field theory}

The Galilei Lie algebra (here for the case $d=1$ only)
is generated from the set ${\cal G}=\{X_{-1},Y_{-1/2},Y_{1/2},\}\subset
{\cal S}_{\rm fin}$.
Thus any Schr\"odinger invariant theory will have to satisfy
the constraints following from Galilean invariance as well. These conditions
are well known for a long time \cite{Levy67} and we briefly recall the
properites relevant for us. In fact, it is possible to construct a
consistent field theory which is Galilei-covariant from the following
postulates. The simplest example of this is second-quantized ordinary
non-relativistic quantum mechanics.

States are rays in a Hilbert space and the dynamical variables
are operators in the Hilbert space, with the usual rules for the calculation
of probabilities. The Galilei group acts by a unitary projective
representation ${\cal U}(g)$ in the Hilbert space.
If $\phi(r,t)$ is a field of the theory, it is required
to transform locally
\BEQ \label{eq:LocTra}
{\cal U}(g)^{-1} \phi(r,t) {\cal U}(g) =
\exp\left[ \frac{i m}{2} ( {v}^2 t + 2{vr})\right]
\phi(r+vt+a,t+\beta)
\EEQ
States are characterized by the Casimir operators, whose eigenvalues
are mass and spin (if $d>1$).
Since we deal with projective representations, we obtain a unitary
representation of a central extension of $\cal G$ by a one-dimensional
Lie algebra generated by $M_0$, see eq.~(\ref{eq:SCHComm}).
This implies, since the extension is
non-trivial because $\cal G$ is non-semisimple,
that a physically trivial transformation may result in a
modification of the phase of the state vector which depends on the mass
of the system. Galilei covariance thus
requires the Bargmann superselection rule of the mass \cite{Barg54}
which states that for an interaction of particles of the form
\BEQ
A + B + \cdots \rar A' + B' + \cdots
\EEQ
one must have
\BEQ \label{eq:Barg}
m_A + m_B + \cdots = m_{A'} + m_{B'} + \cdots
\EEQ
This implies that no Galilean field can be hermitian unless it it massless.
We see that the mass plays quite a distinct role in non-relativistic theories
as compared to relativistic ones. We emphasize that the mass no longer
describes the deviation from critical behaviour in our context, as it does
in relativistic theories. Masses should, in the light of (\ref{eq:Barg}),
rather be regarded as some kind of analogon of a charge \cite{Levy67}.
Non-vanishing correlations are of the type
\BEQ
\langle \phi_a (\vec{r},t) \phi_b^* (\vec{r}\,',t') \rangle \sim
\delta_{m_a , m_b} \, {\cal X}_{a,b}(\vec{r}-\vec{r}\,',t-t')
\EEQ
and similarly for higher order correlations. This also holds by analytic
continuation for Euclidean theories. We shall rederive the superselection rule
eq.~(\ref{eq:Barg}) in several cases below.

A further remark is in order here. In general, in the context of a statistical
system, the mass will contain non-universal factors which merely serve to
define the time scale. Here we are interested in the {\em ratios} of masses
of different scaling fields, which are universal.

While the Bargmann superselection rule provides a restriction not present in
relativistic theories, Galilean field theory is considerably less restricted
in many other aspects. For example, the consequences of locality in
Galilean field theory are no longer
sufficient to prove neither the CPT nor the spin-statistics theorem, see
\cite{Levy67} for a full discussion.

\subsection{Dynamical scaling}

Finally, we recall some facts about response functions and dynamical
scaling, following \cite{Hohe77,Kreu81}.
Consider a scaling field $\phi(\vec{r},t)$. In systems described by
a Hamiltonian one can define its conjugate field $h(\vec{r},t)$. Then the
linear response function $\chi(\vec{k},\omega)$ is defined in
momentum-frequency space by
\BEQ
<\phi(\vec{k},\omega)> = \chi(\vec{k},\omega) h(\vec{k},\omega)
\EEQ
where the field $h$ is taken to be infinitesimal, the average is determined
from the time-dependent probability distribution in the presence of $h$ and the
system is assumed to start from equilibrium at $t\rar-\infty$.
The Fourier transforms are
\BEA
h(\vec{r},t) &=& \int \frac{d\vec{k}}{(2\pi)^{d}} \int_{-\infty}^{\infty}
\frac{d\omega}{2\pi} e^{i(\vec{k}\cdot\vec{r}-\omega t)} h(\vec{k},t)
\label{eq:Fou1}\\
\chi(\vec{k},\omega ) &=& \int_{0}^{\infty} dt \int d\vec{r}
e^{i(\omega t -\vec{k}\cdot \vec{r})} G(\vec{r},t) \label{eq:Fou2}
\EEA
We are interested in time-delayed correlation functions
\BEQ \label{eq:PhiCorr}
C_{\phi}(\vec{r},t) = < \phi(\vec{r},t)\phi(\vec{0},0)>_{h=0}
-<\phi(\vec{r},t)>_{h=0}<\phi(\vec{0},0)>_{h=0}
\EEQ
and define its Fourier transform $C_{\phi}(\vec{k},\omega )$
according to eq.~(\ref{eq:Fou1}) and the
equal-time correlation function $C_{\phi}(\vec{k})$ as the spatial Fourier
transform of (\ref{eq:PhiCorr}) with $t=0$ or
\BEQ
C_{\phi}(\vec{k}) = \int_{-\infty}^{\infty} \frac{d\omega}{2\pi}
C_{\phi}(\vec{k},\omega )
\EEQ
{}From causility, it can be shown (see \cite{Kreu81}) that the response
function $\chi(\vec{k},\omega)$ is an analytic function of the complex
frequency $\omega$ in the upper half-plane and its real and imaginary
parts satisfy the Kramers-Kronig dispersion relations. For classical
systems with a Hamiltonian, the fluctuation-dissipation theorem states
\cite{Hohe77,Kreu81}
\BEQ
C_{\phi}(\vec{k},\omega) = \frac{2 k_{B} T}{\omega} \Im \chi(\vec{k},\omega)
\EEQ
where $\Im$ denotes the imaginary part.
The hypothesis of dynamical scaling now asserts that at a static critical point
\cite{Ferr67,Hohe77}
\BEQ
\chi(\vec{k},\omega) = {\cal A}\, k^{2x-d} \Phi( {\cal B}\omega k^{-\theta})
\EEQ
in the scaling limit $\omega\rar 0, k\rar 0$ with $\omega k^{-\theta}$ fixed,
where $\theta=z$ is the dynamical (anisotropy) exponent,
$x$ a scaling dimension,
$\Phi$ is a universal scaling function and
${\cal A},{\cal B}$ are non-universal constants.

\section{Multipoint correlations from Schr\"odinger invariance}

We now turn to derive the consequences of Schr\"odinger invariance for
the correlations. In general, we expect a scaling field $\phi(\vec{r},t)$ to be
characterized by its mass $\cal M$, its scaling dimension $x$ and its
spin $s$ (which we take to be zero throughout, but see \cite{Hage72} for the
case of spin $1/2$). The discussion will be exclusively for $d=1$, but the
extension to arbitrary $d$ is immediate. The transformation of
$\phi(r,t)$ will contain terms describing the space-time coordinate
change given by $\lmb(r,t)$, the scaling as described by the Jacobian
of $\lmb(r,t)$ and the change in the phase which is a peculiar feature of
non-relativistic systems. Under infinitesimal coordinate changes,
we have the transformations
\BEA
[ X_n , \phi (r,t)] &=& \left( t^{n+1} \partial_t + \frac{n+1}{2} t^n
r\partial_r +\frac{n(n+1)}{4} {\cal M} t^{n-1} r^2 + (n+1)\frac{x}{2} t^n
\right) \phi(r,t) \nonumber \\
{} [ Y_m , \phi(r,t)] &=& \left( t^{m+1/2}\partial_r + \left( m+\frac{1}{2}
\right) {\cal M} t^{m-1/2} r \right) \phi(r,t) \label{eq:CovT}
\EEA
Taking over the conformal terminology of \cite{Bela84},
we call a field {\em primary},
if it satisfies (\ref{eq:CovT}) for all $n$ integer and all $m$ half-integer.
A field is called {\em quasi-primary}, if it satisfies (\ref{eq:CovT})
for the finite-dimensional subalgebra ${\cal S}_{\rm fin}$ only. Consider
multipoint correlators
$\langle \phi_a (r_a, t_a) \phi_b (r_b, t_b) \ldots \phi_{y}^* (r_y,t_y)
\phi_{z}^* (r_z,t_z)\rangle$
of quasiprimary fields and we derive the restrictions following from the
hypothesis of their covariant transformation under ${\cal S}_{\rm fin}$.
We shall use the short-hand
\BEQ
\partial_a = \frac{\partial}{\partial t_a} \;\; ; \;\;
D_a = \frac{\partial}{\partial r_a}
\EEQ

We shall not consider explicitly the action of the generator $M_0$, because
invariance with respect to it follows from the Bargmann superselection rule.

\subsection{Two-point function in the bulk}

We consider the two-point function
\BEQ
F = F(r_a, r_b ; t_a , t_b) =
\langle \phi_a(r_a, t_a) \phi_b^* (r_b, t_b) \rangle
\EEQ
of quasiprimary fields $\phi_{a,b}$ in the infinite geometry in both time and
space. Invariance under translations in time and space implies
$F=F(r,\tau)$, where $r=r_a -r_b$, $\tau=t_a-t_b$. Invariance under scale
transformations generated by $X_0$ requires
\BEQ
\left( t_a \partial_a + \frac{1}{2} r_a D_a + \frac{x_a}{2} +
t_b \partial_b + \frac{1}{2} r_b D_b + \frac{x_b}{2} \right) F(r,\tau) = 0
\EEQ
which is rewitten as, with $x=\frac{1}{2} (x_a + x_b)$
\BEQ
\left( \tau \partial_{\tau} + \frac{1}{2}
r \partial_{r} + x \right) F (r,\tau) = 0
\EEQ
We write the solution in the form
\BEQ \label{eq:ScalSol}
F(r,\tau) = \tau^{-x} G \left( \frac{r^2}{\tau} \right)
\EEQ
which is nothing but the scaling form eq.~(\ref{eq:ScalForm}). New information
comes from requiring Galilei invariance ($Y_{1/2}$)
\BEA
&& \left( t_a D_a + {\cal M}_a r_a + t_b D_b - {\cal M}_b r_b \right) F(r,\tau)
\nonumber \\
&=& \left( \tau \partial_r +
{\cal M}_a r_a-{\cal M}_b r_b \right) F(r,\tau) = 0
\EEA
and we obtain two conditions
\BEQ
\left\{ \begin{array}{r} {\cal M}_a - {\cal M}_b = 0 \\
\left( \tau \partial_r + {\cal M}_a r \right) F(r,\tau) = 0
\end{array} \right.
\EEQ
We recognize in the first of these the
Bargmann superselection rule eq.~(\ref{eq:Barg}).
Combining with scale invariance (\ref{eq:ScalSol}) we
find
\BEQ
G(u) = G_0 \exp \left( -\frac{{\cal M}_a}{2} u \right)
\EEQ
We remark that the form of $F(r,\tau)$ in (\ref{eq:ScalSol}) is not
an arbitrary ansatz. This
can be seen by first solving the conditon of Galilei invariance before using
scale invariance.

Finally, invariance under the special Schr\"odinger transformation $X_1$ gives
\BEQ
\left( t_{a}^{2}\partial_a + t_a r_a D_a + {\cal M}_a r_{a}^{2} + x_a t_a +
t_{b}^{2}\partial_b + t_b r_b D_b - {\cal M}_b r_{b}^{2} +
x_b t_b \right) F(r,\tau) = 0
\EEQ
which is seen as before to lead to the conditions
\BEQ
\left\{ \begin{array}{r}
x = x_a = x_b \\
{\cal M}_a - {\cal M}_b = 0 \\
G' + \frac{1}{2}{\cal M}_a G = 0
\end{array} \right.
\EEQ
where the prime denotes the derivative.
The last two of these had already been obtained before. The final
result is, where $\Phi_0$ is a normalization constant
\BEQ \label{ZweiPunkt}
F = \delta_{x_a,x_b} \delta_{{\cal M}_a , {\cal M}_b} \Phi_0
( t_a - t_b )^{-x_a} \exp \left[ - \frac{{\cal M}_a}{2}
\frac{(r_a - r_b)^2}{t_a - t_b} \right]
\EEQ
which should be understood in the scaling limit. This had been announced before
\cite{Henk92} for the special case of equal masses. Since for a non-conserved
order parameter van Hove theory leads to a dynamic exponent $z=2$, it is
not surprising that we recover in this case the form eq.~(\ref{eq:CRF})
as found for $d=2$ by conformal invariance \cite{Card85}.

\subsection{Three-point function in the bulk}

Consider the three-point function
\BEQ
F = F(r_a,r_b,r_c; t_a,t_b,t_c) =
\langle \phi_a (r_a,t_a) \phi_b (r_b,t_b) \phi_c^* (r_c,t_c) \rangle
\EEQ
Translation invariance in both space and time let $F$ only depend on distances,
$F=F(r,s; \tau,\sig)$ where
\BEQ
r = r_a - r_c \;\; , \;\; s = r_b - r_c \;\; , \;\;
\tau = t_a - t_c \;\; , \;\; \sig = t_b -t_c
\EEQ
Scale invariance requires that
\BEA
&& \sum_{i=a}^{c} \left( t_i \partial_i + \frac{1}{2} r_i D_i + \frac{1}{2} x_i
\right) F (r,s;\tau,\sig) \nonumber \\
&=& \left( \tau\partial_{\tau} + \sig\partial_{\sig} + \frac{1}{2} r\partial_r
+\frac{1}{2} s\partial_s +
\frac{1}{2}\left( x_a + x_b + x_c \right) \right) F = 0
\EEA
Making the ansatz
\BEQ \label{eq:AnsatzG}
F(r,s;\tau,\sig) = \tau^{-\rho_1} \sig^{-\rho_2} (\tau-\sig)^{-\rho_3}
G(r,s;\tau,\sig)
\EEQ
as motivated by the corresponding result for the three-point function as
obtained from conformal invariance \cite{Poly70}, we find
\BEQ \label{eq:ScalI}
\left\{ \begin{array}{c}
\rho_1 + \rho_2 +\rho_3 = \frac{1}{2} ( x_a + x_b + x_c ) \\
\left( \tau\partial_{\tau} + \sig\partial_{\sig} + \frac{1}{2} r\partial_r
+\frac{1}{2} s\partial_s \right) G(r,s;\tau,\sig) = 0
\end{array} \right.
\EEQ
Note that the relation found between the exponents $\rho_i$ and the scaling
dimensions $x_i$ must be satisfied for any scale-invariant system.
{}From Galilei invariance we get, with $\eps_a=\eps_b=-\eps_c=1$
because of eq.~(\ref{eq:LocTra})
\BEA
&& \sum_{i=a}^{c}
\left( t_i D_i + \eps_i {\cal M}_i r_i \right) F(r,s;\tau,\sig) \\
&=& \left( \tau \partial_{r} +
\sig\partial_{s} + {\cal M}_a r + {\cal M}_b s
+ \left( {\cal M}_a + {\cal M}_b - {\cal M}_c \right) r_c \right)
F(r,s;\tau,\sig) = 0 \nonumber
\EEA
which leads to the conditions
\BEQ \label{eq:ScalG}
\left\{ \begin{array}{c}
{\cal M}_a + {\cal M}_b - {\cal M}_c = 0 \\
\left( \tau \partial_{r} +\sig\partial_{s}
+ {\cal M}_a r + {\cal M}_b s\right) G(r,s;\tau,\sig) = 0
\end{array} \right.
\EEQ
and we recognize again the Bargmann superselection
rule for the masses. The equation
for $G$ can be further simplified by setting
\BEQ \label{eq:AnsatzH}
G(r,s;\tau,\sig) = \exp \left( -\frac{{\cal M}_a}{2} \frac{r^2}{\tau}
-\frac{{\cal M}_b}{2} \frac{s^2}{\sig} \right) H(r,s;\tau,\sig)
\EEQ
Since the first factor is scale invariant,
$H$ satisfies the same eq.~(\ref{eq:ScalI})
as does $G$, but the second eq.~(\ref{eq:ScalG}) becomes
\BEQ
\left( \tau \partial_{r} +\sig\partial_{s} \right) H(r,s;\tau,\sig) = 0
\EEQ
Finally, invariance under the special Schr\"odinger transformation requires
\BEQ
\sum_{i=a}^{c} \left( t_{i}^{2} \partial_i + t_i r_i D_i + \frac{1}{2}
\eps_i {\cal M}_i r_{i}^{2} + t_i x_i \right) F(r,s;\tau,\sig) = 0
\EEQ
First, we use the form (\ref{eq:AnsatzG}). Then we get
$({\cal T} + {\cal D})G = 0$, where $\cal D$ is a differential operator and
\BEQ
{\cal T} = t_a (-\rho_1 -\rho_3 + x_a ) + t_b (-\rho_2 -\rho_3 + x_b ) +
t_c (-\rho_1 -\rho_2 + x_c )
\EEQ
The requirement that $\cal T$ vanishes leads together with scale invariance
(\ref{eq:ScalI}) to
\BEQ
\rho_2 = \frac{1}{2} (x_b +x_c -x_a) \;\; , \;\;
\rho_1 = \frac{1}{2} (x_a +x_c -x_b) \;\; , \;\;
\rho_3 = \frac{1}{2} (x_a +x_b -x_c)
\EEQ
Secondly, we use eq.~(\ref{eq:AnsatzH}) and get $({\cal T}' + {\cal D}') H=0$
where
\BEQ
{\cal T}' = \frac{1}{2} r_c^2
\left( {\cal M}_a + {\cal M}_b - {\cal M}_c \right)
\EEQ
This vanishes due to the Bargmann superselection rule.
The differential operator
${\cal D}'$ finally takes the form where we used that $H$ is scale as well as
Galilei invariant
\BEQ
\left( \tau^2 \partial_{\tau} +\sig^2 \partial_{\sig}
+\tau r\partial_r + \sig s\partial_s \right) H(r,s;\tau,\sig) = 0
\EEQ
It remains to solve the resulting system of
first order linear partial differential
equations. This is done in Appendix~A, with the result
\BEQ \label{eq:HLoes}
H(r,s;\tau,\sig) = \Psi
\left( \frac{(r\sig -s\tau)^2}{(\sig-\tau)\sig\tau} \right)
\EEQ
where $\Psi =\Psi_{ab,c}$ is an arbitrary function.
The final result is, with the appropriate scaling limits understood
\BEA
\lefteqn{ F = \delta_{{\cal M}_a+{\cal M}_b,{\cal M}_c}
(t_a -t_c)^{-\frac{1}{2}(x_a +x_c -x_b)}
(t_b -t_c)^{-\frac{1}{2}(x_b +x_c -x_a)}
(t_a -t_b)^{-\frac{1}{2}(x_a +x_b -x_c)} } \nonumber \\
&\times& \exp\left[ -\frac{{\cal M}_a}{2}\frac{(r_a -r_c)^2}{t_a -t_c}
-\frac{{\cal M}_b}{2}\frac{(r_b -r_c)^2}{t_b -t_c} \right]
\Psi \left(\frac{[(r_a-r_c)(t_b-t_c)-(r_b-r_c)(t_a-t_c)]^2}{(t_a-t_b)
(t_a-t_c)(t_b-t_c)} \right) \nonumber \\
&& ~ \label{DreiPunkt}
\EEA
In particular, it follows that any three-point function $<\phi\phi\phi^*>$ of
a massive field with itself vanishes.

The results eqs.~(\ref{ZweiPunkt},\ref{DreiPunkt}) deserve some comments.
It is instructive to compare them with
those for the two-point and the three-point functions obtained from the
requirement of conformal invariance by Polyakov \cite{Poly70}, see also
\cite{Card90,Osbo93}. Conformal
invariance completely specifies the form of the two- and three-point functions
in any number of space dimensions. Also, two-point correlations of quasiprimary
fields must vanish if the scaling dimensions are different. We reproduce this
result in eq.~(\ref{ZweiPunkt}), but add the stronger requirement of the
Bargmann mass selection rule. The exponential behaviour of the
two-point scaling function is a consequence of Galilei invariance, while
$\Phi_0$ is merely a normalization. Our results do depend on the explicit
realization of the Galilei transformation as given by the generator $Y_{1/2}$.
If we had considered the Schr\"odinger group with a potential present, the
realization of the generators is different
\cite{Nied74} and we would have found
a different form of the correlations. Turning to the three-point function
eq.~(\ref{DreiPunkt}), we
observe that the purely time-dependent factors reproduce the familar form
of the conformal three-point function which is completely symmetric in the
times $t_a,t_b,t_c$. We then note that this symmetry is not met by the
exponential factors, determined from Galilei invariance and we also note
the appearance of the Bargmann mass selection rule, not present in conformal
invariance. Generalization of our results to higher space dimension $d$
merely requires to check for rotation invariance, which is obviously satisfied.

\subsection{Two-point function in semi-infinite space}

Having studied correlations of quasiprimary fields in the infinite geometry,
we now consider the effect of surfaces. Consider
a free surface at $r=0$. It is kept
invariant under the transformations generated by the subalgebra ${\cal S}_X$,
but space translations and Galilei transformations will no longer
leave the system invariant. Nevertheless, it is known that conformal invariance
can be used in analogous situations to constrain the two-point correlation
function \cite{Card84}. For quasiprimary fields, we require covariance only
under the subalgebra ${\cal S}_X \cap {\cal S}_{\rm fin}
=\{ X_{-1},X_0,X_{1}\}$.

Consider the two-point function of quasiprimary fields
\BEQ
F = F(r_a,r_b;t_a,t_b) =\langle \phi_a (r_a,t_a) \phi_b^* (r_b,t_b) \rangle
\EEQ
and we require space points to be in the right half-plane, i.e.
$r_a, r_b \geq 0$.
Time translation invariance gives $F=F(r_a,r_b;\tau)$, with $\tau=t_a-t_b$.
{}From scale invariance we obtain
\BEA
&& \sum_{i=a}^{b} \left( t_i \partial_i +\frac{1}{2} r_i D_i +\frac{1}{2} x_i
\right) F \nonumber \\
&=& \left( \tau \partial_{\tau} + \frac{1}{2} r_a D_a + \frac{1}{2} r_b D_b
+ x \right) F = 0
\EEA
where $x = \frac{1}{2}(x_a + x_b)$. On the other hand, from the invariance
under the special Schr\"odinger transformation we have, with $\eps_a=-\eps_b=1$
\BEA
&& \sum_{i=a}^{b} \left( t_{i}^{2} \partial_i + t_i r_i D_i
+ \frac{1}{2}\eps_i {\cal M}_i r_{i}^2 + t_i x_i \right) F \nonumber \\
&=& \left( \tau^2 \partial_{\tau} +\tau r_a D_a
+t_b \left( 2\tau \partial_{\tau} +r_a D_a + r_b D_b \right)
\frac{ }{} \right. \nonumber \\
&& \left. + \frac{1}{2}{\cal M}_a r_{a}^2 - \frac{1}{2}{\cal M}_b r_{b}^2
+ t_a x_a + t_b x_b \right) F \nonumber \\
&=& \left( \tau^2 \partial_{\tau} +\tau r_a D_a
+ \frac{1}{2}{\cal M}_a r_{a}^2 - \frac{1}{2}{\cal M}_b r_{b}^2
+ \tau x_a \right) F = 0 \label{eq:SSurf}
\EEA
where in the last equation the scale invariance of $F$ was used. Now, we make
the ansatz
\BEQ
F(r_a,r_b;\tau) = \tau^{-x} G(u,v) \;\; , \;\; u =\frac{r_{a}^2}{\tau}
\;\; , \;\; v =\frac{r_{b}^2}{\tau}
\EEQ
which solves for scale invariance, while eq.~(\ref{eq:SSurf}) gives
\BEQ
\left\{ \begin{array}{c}
x = x_a = x_b \\
\left( u\partial_u - v\partial_v + \frac{1}{2}{\cal M}_a u -
\frac{1}{2}{\cal M}_b v \right) G(u,v) = 0
\end{array} \right.
\EEQ
The general solution of this is found using the method of
characteristics \cite{Kamk59}
\BEQ
G(u,v) = \chi( u v ) \exp\left[ -\frac{1}{2}{\cal M}_a u -
\frac{1}{2}{\cal M}_b v \right]
\EEQ
where $\chi$ is an arbitrary function. The final result is
\BEQ \label{SOber}
F = \delta_{x_a,x_b} (t_a-t_b)^{-x_a} \chi\left(\frac{r_a r_b}{t_a -t_b}\right)
\exp\left[ -\frac{{\cal M}_a}{2}\frac{r_{a}^2}{t_a-t_b}
-\frac{{\cal M}_b}{2}\frac{r_{b}^2}{t_a-t_b} \right]
\EEQ
We note that analogously to the conformal result \cite{Card84}, the scaling
dimensions have to agree, while in this case we do not have a constraint
on the masses ${\cal M}_{a,b}$, since the system is {\em not}
Galilei-invariant.

The function $\chi$ is partially determined from a few consistency conditions.
We should expect to recover the bulk behaviour for large distances to
the surfaces, that is for $r_a r_b /\tau \rar \infty$. Therefore, up
to normalization
\BEQ \label{eq:Con1}
\chi(u) \simeq \delta_{{\cal M}_a , {\cal M}_b} e^{{\cal M}_a u}
\;\; ; \;\; u \rar \infty
\EEQ
On the other hand, for a free surface where the field vanishes, we expect
absence of any correlations and thus
\BEQ \label{eq:Con2}
\chi(0) = 0
\EEQ

Indeed, this is exactly the behaviour obtained from the method of images.
We have for the surface correlation $G_s$ in terms of bulk correlations $G_b$
\BEA
G_s (r,r';\tau) &=& G_b (r-r',\tau) - G_b (r+r',\tau) \nonumber \\
&=& G_0 \tau^{-x}
\left( \exp\left[-\frac{{\cal M}}{2}\frac{(r-r')^2}{\tau}\right]
- \exp\left[-\frac{{\cal M}}{2}\frac{(r+r')^2}{\tau}\right] \right)
\nonumber \\
&=& 2 G_0 \tau^{-x} \sinh\left({\cal M}\frac{r r'}{\tau}\right)
\exp\left[-\frac{{\cal M}}{2}\frac{r^2+{r'}^2}{\tau}\right]
\EEA
and we identify $\chi(u)=2 G_0\sinh({\cal M} u)$
in agreement with the consistency
conditions eqs.~(\ref{eq:Con1},\ref{eq:Con2}).

Finally, if we were to impose in addition translation invariance in space,
invariance under Galilei transformations is also implied and we do recover
the bulk result (\ref{ZweiPunkt}) for the two-point function.

\subsection{Two-point function for a non-stationary state}

We now consider a situation with a boundary condition at a fixed time.
Boundary conditions of this type will be kept invariant by the subalgebra
${\cal S}_Y$ together with the scale transformation $X_0$, or rather
its finite-dimensional subalgebra
$\{ X_0,Y_{-1/2},Y_{1/2}\}$ for quasiprimary fields.
For example, this may correspond to the situation when a system is
in a predefined initial state and relaxes towards its critical
equilibrium state \cite{Jans89}. Consider
the two-point function of quasiprimary fields
\BEQ
F = F(r_a,r_b;t_a,t_b) =\langle \phi_a (r_a,t_a) \phi_b^* (r_b,t_b) \rangle
\EEQ
Invariance under space translations implies $F=F(r;t_a,t_b)$ with
$r=r_a -r_b$. We next demand invariance under Galilei transformations,
with $\eps_a=-\eps_b=1$
\BEA
&& \sum_{i=a}^{b} \left( t_i D_i + \eps_i {\cal M}_i r_i \right) F \nonumber \\
&=& \left( (t_a-t_b)\partial_r +{\cal M}_a r_a -{\cal M}_b r_b \right) F =0
\EEA
In analogy to what was done before, this implies the Bargmann superselection
rule \linebreak ${\cal M}_a = {\cal M}_b$ and
\BEQ \label{eq:TSurf}
F(r;t_a,t_b) = G(t_a,t_b) \exp\left(
-\frac{{\cal M}_a}{2} \frac{r^2}{t_a -t_b}\right)
\EEQ
Scale invariance demands that
\BEQ
\sum_{i=a}^{b} \left( t_i \partial_i +\frac{1}{2} r_i D_i + \frac{1}{2} x_i
\right) F = 0
\EEQ
Inserting (\ref{eq:TSurf}), we find
\BEQ
G(t_a,t_b) = t_{a}^{-(x_a+x_b)/2} \Phi( t_a / t_b)
\EEQ
where $\Phi(v)$ is an undetermined function. The final result is
\BEQ \label{TOber}
F = \delta_{{\cal M}_a , {\cal M}_b}
t_{a}^{-(x_a+x_b)/2}  \Phi( t_a / t_b)
\exp\left( -\frac{{\cal M}_{a}}{2} \frac{(r_a-r_b)^2}{t_a -t_b}\right)
\EEQ
Note that here we have no condition on the exponents because the system is
not invariant under the special Schr\"odinger transformation $X_1$.

A few consistency conditions may be added. If both times $t_a,t_b$ are very
large, we should expect to recover the form eq.~(\ref{ZweiPunkt}) for the
fully infinite space-time. In that case $t_a /t_b \rar 1$ and we should have
\BEQ
\Phi(v) \sim \delta_{x_{a},x_{b}} \left( 1 - v^{-1} \right)^{-(x_a +x_b )/2}
\;\; , \;\; v \rar 1
\EEQ
On the other hand, if $t_b$ is kept finite, one should remain in the
intermediate time scalime regime, termed ``initial critical slip''
in \cite{Jans89}. In this case, one should expect
\BEQ
\Phi(v) \sim v^{\Theta} \;\; , \;\; v \rar \infty
\EEQ
where $\Theta$ is related to the independent ``slip exponent'' defined in
\cite{Jans89}. We shall present an example of this in section~4.

\subsection{Ward identity}

We now consider the effect of arbitrary coordinate transformations on
correlations. We suppose that the system under consideration is
described by a local action in $d+1$ dimensions. This is obviously satisfied
for static, but strongly anisotropic systems with local interactions.
For many dynamical problems in $d$ space dimensions
which are at first defined via their equation of motion, there exists
an equivalent equilibrium problem in $d+1$
dimensions, usually supplemented with
a ``disorder conditions'' to maintain the strong
anisotropy \cite{Ruja87,Kand90}.
Then, considering the change of the action induced by an arbitrary coordinate
transformation, the following identity holds, see e.g. \cite{Card90}
\BEA \label{eq:Ward}
&& \sum_{a=1}^{n} \langle \phi_1(\vec{r}_1,t_1)\ldots
\delta\phi_a(\vec{r}_a,t_a)\ldots
\phi_n(\vec{r}_n,t_n) \rangle \\
&+& \int d\vec{R} dT
\langle \phi_1(\vec{r}_1,t_1)\ldots\phi_n(\vec{r}_n,t_n)
T_{ij}(\vec{R},T) \rangle
\partial_i (\delta\vec{r}_{j})(\vec{R},T) = 0 \nonumber
\EEA
where we implicitly assume that in the correlators the Bargmann superselection
rule is satisfied, the time $T$ is denoted as the
zeroth component of the coordinate $\vec{R}$
and $T_{ij}$ is the stress-energy tensor. As eq.~(\ref{eq:Ward}) is written,
we assume $\delta \phi$ to contain all
variations of the field $\phi$ (including
changes of its phase) and $T_{ij}$ to describe
all changes of the action
used to calculate the averages $\langle \ldots \rangle$. The discussion
presented here remains at the formal level of the equations of motion.
We discard the possibility of anomalies which may arise from renormalization
effects. Detailed discussions of these are available for conformal invariant
theories, see \cite{Bela84,Card90}, but for Schr\"odinger invariance,
the analogous developments have not yet been done.

Since correlations are supposed to be invariant under infinitesimal
Schr\"odinger transformations, we obtain
a few constraints on the stress-energy
tensor in complete analogy with the corresponding results
for conformal invariance
\cite{Card90}. The form of the $\delta \vec{r}_j$ is taken from the
generators eq.~(\ref{eq:SGen}). Rotation invariance implies that
$T_{ij}$ is symmetric in space
\BEQ
T_{ij} = T_{ji} \;\; , \;\; i,j=1,\ldots,d
\EEQ
Scale invariance gives the ``trace condition'' (here written in Euclidean form)
\BEQ \label{eq:TraCond}
2 T_{00} + \sum_{i=1}^{d} T_{ii} = 0
\EEQ
which is the analogue of the vanishing trace condition in conformal
invariance \cite{Card90} and the factor $2$ comes from $\theta=2$.
Eq.~(\ref{eq:TraCond}) is satisfied in
free non-relativistic field
theory \cite{Hage72} (where it is written in Minkowskian form).
Interacting non-relativistic field theories may give
rise to anomalies, see \cite{Berg92}. From Galilei invariance we find
\BEQ
T_{0i} = 0 \;\; , \;\; i=1,\ldots,d
\EEQ
The requirement of special Schr\"odinger invariance does not add any further
condition on $T_{ij}$. We have thus seen that
\BEQ
\left. \begin{array}{l}
\mbox{\rm translation invariance in space and time} \\
\mbox{\rm rotation invariance in space} \\
\mbox{\rm anisotropic scale invariance with $\theta=2$} \\
\mbox{\rm Galilei invariance} \\
\mbox{\rm local interactions} \end{array} \right\}
\Longrightarrow \mbox{\rm Schr\"odinger invariance}
\EEQ
in analogy to the conformal result \cite{Card90}. In fact, using formally
the equations of motion, we may even verify invariance under the entire
infinite algebra $\cal S$. This is for the time being the only indication
that the generalization beyond ${\cal S}_{\rm fin}$ might be sensible. Again,
the same type of result also holds for conformal invariance (when the central
charge vanishes).

We do not go into a discussion of the possible anomaly structure here. As a
preliminary exercise to that, we show in Appendix~B that the Schr\"odinger
algebra eq.~(\ref{eq:SCHComm}) does not admit any non-conventional central
extension besides the familiar Virasoro form for the generators $X_n$.

\subsection{Summary}

The main results of this section are the explicit expressions for the
two- or three-point Schr\"odinger-covariant
correlations in either an infinite or a semi-infinite
geometry as given in
eqs.~(\ref{ZweiPunkt},\ref{DreiPunkt},\ref{SOber},\ref{TOber}). Although
the general form is quite similar to the corresponding results found
from conformal invariance \cite{Poly70,Osbo93,Card84}, there
are some properties
which come from the non-relativistic nature of the symmetry. The first
one is the Bargmann superselection rule \cite{Barg54} for the masses.
Secondly, the
space dimension $d$ has mainly the role of a parameter, at least for
the quasiprimary fields only considered here, whereas the non-trivial
group structure capable  of extension to an infinite-dimensional algebra
only occurs in the ``time'' coordinate.

We remark that there is a certain analogy between Schr\"odinger invariance
and conformal invariance close to a free surface \cite{Card84}. In both
cases, the pair of {\em complex} linear projective transformations
characteristic for full conformal invariance gets replaced by the subgroup
of a single {\em real} linear projective transformation.

The results obtained only use the finite-dimensional algebra. It remains
an open problem how to extend Schr\"odinger invariance to the full
infinite-dimensional algebra and find the scaling functions.

\section{Tests of Schr\"odinger invariance in exactly solvable models}

We now turn to test the predictions obtained in the last section
for some correlations in the context of some exactly solvable strongly
anisotropic critical systems. We do not include here among the tests
the well-known
fact that the Green's functions of the free Schr\"odinger equation and
the diffusion equation reproduce eq.~(\ref{ZweiPunkt}) in any space dimension
with $x=d/2$. We also refrain from discussing the range of possible
applications of the models considered.

\subsection{Kinetic Ising model with Glauber dynamics}

Consider the time-dependent $1D$ Ising model with the classical spin
Hamiltonian ${\cal H} = -J \sum_{i=1}^{L} s_i s_{i+1}$ and $s_i =\pm 1$.
To describe the time dependence, following Glauber \cite{Glau63},
consider the probability distribution function
$P(s_1,\ldots,s_L;t)$. It is often convenient to describe the evolution
of $P$ in terms of a master equation (see \cite{Kreu81,Gunt83} for a detailed
discussion)
\BEA
\lefteqn{ \frac{\partial}{\partial t} P(s_1,\ldots,s_L;t) =} \\
&-& \left( \sum_i w_i (s_i) \right) P(s_1,\ldots,s_L;t) +
\sum_i w_i(-s_i) P(s_1,\ldots,-s_i,\ldots,s_l;t) \nonumber
\EEA
where the $w_i$ are the rates describing the transitions between spin
configurations. The following consistency conditions have to be kept.
The first is probability conservation when summing over all configurations
$\{s\} = (s_1,\ldots,s_L)$
\BEQ
\sum_{\{s\}} P(\{s\};t) = 1
\EEQ
to be kept at all times. Second, the equilibrium distribution
$P_{\rm eq} \sim e^{-\beta{\cal  H}}$ has to be a stationary solution
of the master equation, where $\beta$ is the inverse temperature.
This is usually implemented via detailed balance
\BEQ
\frac{w_i(-s_i)}{w_i(s_i)} =
\frac{\exp[-\beta J s_i(s_{i-1}+s_{i+1})]}{\exp[\beta J s_i(s_{i-1}+s_{i+1})]}
\EEQ
Finally, averages are obtained from
\BEQ
<X>(t) = \sum_{\{s\}}  X(\{s\}) P(\{s\};t)
\EEQ

Glauber \cite{Glau63} showed that the particular choice
\BEQ
w_i(s_i) = \frac{\alpha}{2} \left( 1 -\frac{\gamma}{2} s_i (s_{i-1}+s_{i+1})
\right)
\EEQ
where $\alpha$ is the constant transition rate and $\gamma=\tanh(2\beta J)$,
renders the model completely integrable. We are interested here in his result
for the time-delayed (connected) two-point
function when the system is in thermal
equilibrium at temperature $T$ \cite{Glau63}
\BEQ
G(r_1-r_2,t_1-t_2)=<s_{r_1}(t_1) s_{r_2}(t_2)>_c = e^{-\alpha t}
\sum_{\ell} \eta^{|r-\ell |} I_{\ell}(\alpha \gamma t)
\EEQ
where $r=r_1-r_2$, $t=t_1-t_2$, $\eta=\tanh(\beta J)$, $I_{\ell}$ is a modified
Bessel function and the sum extends over the whole lattice. If there are no
correlations between spins in the initial state, only the term with
$\ell=r$ would be present.

To analyse this, recall the asymptotic form, as $x\rar\infty$ \cite{Sing83}
\BEQ
I_{\ell}(x) \simeq (2\pi x)^{-1/2} \exp\left( x -\frac{\ell^2}{2 x}\right)
\left( 1 + {\cal O}(x^{-1})\right)
\EEQ
and obtain in the scaling limit
$r\rar\infty, t\rar\infty$ with $u=r^2 /t$ fixed
\BEA
G(r,t) &\simeq& e^{-(1-\gamma)\alpha t} (2\pi \gamma \alpha t)^{-1/2}
\left\{ \exp\left(-\frac{r^2}{2\gamma\alpha t}\right)
+\sum_{\ell\neq 0} \eta^{|\ell |}
\exp\left(-\frac{(r+\ell)^2}{2\gamma\alpha t}\right) \right\} \nonumber \\
&\sim& (2\pi \alpha t)^{-1/2}
\exp\left(-\frac{r^2}{2\alpha t}\right)
\EEA
where in the last equation we performed the zero-temperature limit (since the
$1D$ Ising model has its critical point at $T=0$). This holds exactly for
vanishing correlations in the initial
state and up to scaling corrections otherwise.
This is indeed in agreement with the predicted two-point function
eq.~(\ref{ZweiPunkt}) and we identify $x=1/2$ and ${\cal M}=1/(2\alpha)$.

\subsection{Lattice diffusion with exclusion}

Consider a system of many particles performing random walks on a chain
of $L$ sites. Each site can be either empty or occupied. The dynamics is
defined as follows \cite{Kand90}. First,
pair all neighboring sites which can be done in two ways to be labelled
$\cal A$, chosen at odd times and $\cal B$ which is chosen at even
times. At every time step, the dynamics of each pair is
as follows. If both sites of a pair are either  occupied or empty, the state
of the pair is unchanged. If one site is occupied and one empty, the
particle moves to the empty site  with probability $p$ or stays where it
is with probability $1-p$. This stochastic rule for updating is applied
in parallel to all pairs.

The time-delayed particle-particle correlation is
\BEQ
G(r,t) = < n(r,t) n(0,0) >_c
\EEQ
where $n(r,t)=1$ if the site $r$ is occupied at time $t$ and $n(r,t)=0$
otherwise. The average is first over realizations of all possible time
developments and second over an ensemble of initial states that is
stationary with respect to the stochastic process \cite{Kand90}.
$G$ was calculated exactly first for $p=1/2$ \cite{Kand90} and later for
$p$ arbitrary \cite{Schu93} with the result, for even times $t$ and even
sites $r$ (similar results are known for the other cases, see \cite{Schu93})
\BEQ
G(r,t) = \rho (1-\rho) \left[ p^{t}\delta_{-r,t}+\sum_{k=1}^{(t-|r|)/2}
\left(\vekz{(t-r)/2}{k}\right)\left(\vekz{(t-2+r)/2}{k-1}\right)
p^{t-2k}(1-p)^{2k} \right]
\EEQ
where $\rho = N/L$ is the particle density and $N$ is the number of
particles on the lattice. The limit $L\rar\infty$ such that $\rho$ is kept
fixed is understood.
In the scaling limit $r,t\rar\infty$ such that $u=r^2 /t$ is kept fixed,
this simplifies to \cite{Kand90,Schu93}
\BEQ \label{eq:LDiff}
G(r,t) = \rho (1-\rho) (2\pi D t)^{-1/2} \exp\left(-\frac{r^2}{2 D t}
\right)
\EEQ
and the diffusion constant $D=p/(1-p)$. This agrees with the Schr\"odinger
invariance expectation eq.~(\ref{ZweiPunkt}), and we have $x=1/2$ and
${\cal M}=1/D$. Similar results were obtained when quenched weak disorder
is added, and it was shown that if the spatial disorder correlations
decay rapidly enough, the same scaling form (\ref{eq:LDiff}) results with
a modified value of the diffusion constant $D$, see \cite{Schu93} for details.

These results werde derived \cite{Kand90,Schu93} by mapping the system
onto a six-vertex model satisfying a disorder condition. From the transfer
matrix ${\cal T}=e^{-\tau H}$ the quantum Hamiltonian can be obtained.
In the time continuum limit $\tau\rar\infty$ such that $p \tau$
remains constant, the quantum Hamiltonian is found \cite{Schu93} to agree with
the
quantum Hamiltonian obtained directly from the
master equation written in the form $\partial_t P = - H P$, and reads
\cite{Alca93}
\BEQ
H = -\frac{1}{2} \sum_{i=1}^{L} \left[ \sig_{i}^{x}\sig_{i+1}^{x} +
\sig_{i}^{y}\sig_{i+1}^{y} + \Delta \sig_{i}^{z}\sig_{i+1}^{z}
+ (1 - \Delta) \left( \sig_{i}^{z}+\sig_{i+1}^{z}\right) +\Delta -2\right]
\EEQ
where the $\sig^{x,y,z}$ are Pauli matrices and with $\Delta=1$ for lattice
diffusion. We note that this Hamiltonian,
but now with $\Delta=0$, is also obtained \cite{Alca93,Sigg77} from the master
equation of the kinetic Ising model at $T=0$ considered above. The disorder
condition for the vertex model formulation of lattice diffusion above makes the
model undergo a transition of Pokrovsky-Talapov type \cite{Pokr80}. In view of
the common scaling form for both $\Delta=0$ or $1$, and because the spectrum
of $H$ is known to be $\Delta-$ independent \cite{Alca93}, we should expect
this
scaling form to hold independently of the value of $\Delta$.

\subsection{Lifshitz point in the spherical model}

We now consider a static, but strongly an\-iso\-tro\-pic system. The model
is the an\-iso\-tro\-pic next-to nearest neighbor spherical (ANNNS) model,
see \cite{Selk92} and references therein.
Conventionally, it is defined by the Hamiltonian, on a hypercubic lattice
\BEQ \label{eq:RSM}
{\cal H}_{SM} = - \sum_{i,j} J_{\vec{\im}\,\vec{\jm}}\,
\sig_{\vec{\im}}\, \sig_{\vec{\jm}} + \beta^{-1}\zeta
\sum_{i} {\sig_{\vec{\im}}\,}^{2}
\EEQ
where $\sig_i$ are real numbers, the spherical parameter
$\zeta$ is determined from
the constraint $<\sum_i s_i^2 >= {\cal N}$ where $\cal N$ is the number of
sites. Consider the model in $D=d'+d$ dimensions. The couplings
$J_{\vec{\im}\,\vec{\jm}}$ are defined as follows.
First, in all $D$ dimensions, there
is a ferromagnetic nearest neighbor interaction of energy $J>0$. Second,
in the $d$ ``parallel'' directions, there is along the axes an interaction of
energy $\kappa J$ between next-to-nearest neighbors. The Fourier transform of
$J_{\vec{\im}\,\vec{\jm}}$ is
\BEQ
J(\vec{k}\,) = 2 J \left( \sum_{i=1}^{D} \cos k_i + \kappa \sum_{i=1}^{d}
\cos 2k_i \right)
\EEQ
If a phase transition occurs, there
is at $\kappa=\kappa_c=-1/4$ a meeting point
between a paramagnetic, a ferromagnetic and an ordered incommensurate phase,
which is referred to as a Lifshitz point \cite{Horn75}, which is a strongly
anisotropic critical point.
When $d'=1$, one has  $\theta=2$ and those coordinates
in the $d$ ``parallel'' directions will be referred to as ``space'',
while the one remaining direction will be referred to as ``time''.

In order to make the Galilei invariance explicit, we consider here a variant
of this model, which gives the same thermodynamics. The Hamiltonian is
\BEQ
{\cal H} = -\sum_{i,j} \frac{1}{2} J_{\vec{\im}\,\vec{\jm}} \left(
s_{\vec{\im}}^* s_{\vec{\jm}} + s_{\vec{\im}} s_{\vec{\jm}}^* \right)
+\beta^{-1} \zeta \sum_{i} \left| s_{\vec{\im}} \right|^{2}
\EEQ
where now $s_i$ is complex and the spherical constraint is
\BEQ \label{eq:CSpC}
< \sum_i \left| s_{\vec{\im}} \right|^{2} > = 2 {\cal N}
\EEQ
We introduce the Fourier transform of the spin variable
\BEQ
s_{\vec{a}} = (2\pi)^{-D/2} \int d\vec{k}\, \mu_{\vec{k}} \,
e^{i \vec{k}\cdot \vec{a}}
\EEQ
and get
\BEQ
{\cal H} = -\int d\vec{k} \left[J(\vec{k}\,) -\beta^{-1}\zeta \right]
\left| \mu_{\vec{k}}\right|^2
= -\int d\vec{k}\, \Lambda(\vec{k}\,) \left| \mu_{\vec{k}}\right|^2
\EEQ
The partition function is ${\cal Z} = \int {\cal D}s\, e^{-\beta{\cal H}}$.
Since
\BEQ
<\mu_{\vec{k}}>=<\mu_{\vec{k}}^2>=0 \;\; , \;\;
<|\mu_{\vec{k}}|^2>=\Lambda^{-1}(\vec{k})
\EEQ
the spherical constraint (\ref{eq:CSpC}) becomes
$\int d\vec{k} \Lambda^{-1}(\vec{k}) = 2{\cal N}$, which is exactly
the same as obtained from ${\cal H}_{SM}$ (see e.g. \cite{Joyc72})
and the free energy is $F=-\beta^{-1} \ln {\cal Z} = 2 F_{SM}$, where
$F_{SM}$ is the free energy obtained from the Hamiltonian ${\cal H}_{SM}$.
Consequently, the model defined by $\cal H$ is in the same universality class
as the one defined by ${\cal H}_{SM}$. In particular, the critical point
is characterized  by the condition $\zeta=\beta J D$. The lower critical
dimension is at $D=3$, the upper critical dimension at $D=7$, provided $d'=1$
which we assume from now on.

Consider the two-point function
$C(\vec{a}-\vec{b}\,)=\Re <s_{\vec{a}}s_{\vec{b}}^*>$ where $\Re$ denotes
the real part and
\BEA
<s_{\vec{a}}s_{\vec{b}}^*> &=& \frac{k_{B} T}{{\cal Z}(2\pi)^D}
\int {\cal D}\mu \int d\vec{k}d\vec{\ell}\,\mu_{\vec{k}}^* \mu_{\vec{\ell}}\,
e^{i(\vec{\ell}\cdot\vec{a}-\vec{k}\cdot\vec{b}\,)}
e^{ -\int d\vec{m}\, \Lambda(\vec{m}\,)
|\mu_{\vec{m}}|^2 }  \nonumber \\
&=& (2\pi)^{-D} k_{B}T\, \int d\vec{k}\, \Lambda^{-1}(\vec{k})\,
e^{i\vec{k}\cdot(\vec{a}-\vec{b}\,)}
\EEA
This has been calculated for integer dimensions $D$ \cite{Frac93}. In the
scaling limit $r\rar\infty, t\rar\infty$ with $r^2 /t$ fixed
($\vec{r}$ and $t$ are the distances in space and time between the
points $a$ and $b$) the result is
\BEA \label{eq:DeuxP}
C(\vec{a}-\vec{b}\,) =
C(\vec{r},t)&=& {\cal A}^2 \,  t^{-(D-3)/2} \Psi\left(\frac{D-3}{4},
\frac{r^2}{2t}\right) \\
{\cal A}^2 &=& \frac{k_B T_c}{J} \sqrt{ \frac{2^{1-D}}{\pi^{2D-3}} }
\left[ \Gamma\left(\frac{1}{4}\right) \right]^{D-2} \nonumber
\EEA
The function $\Psi(a,x)$ is given in table~\ref{TabFun} \cite{Frac93}.
\begin{table}[htb]
\begin{center}
\begin{tabular}{|c|c|c|} \hline
$a$ & $\Psi(a,x)$ & asymptotics \\ \hline
$\frac{1}{4}$ & $x^{1/2} \left[ I_{-1/4}(x/2)+I_{1/4}(x/2)\right] K_{1/4}(x/2)$
& $2 x^{-1/2}$ \\
$\frac{1}{2}$ &
$(\pi^2 x/2)^{1/4}\left[ I_{-1/4}(x) - {\bf L}_{-1/4}(x)\right]$
& $(2/\Gamma(1/4)) x^{-1}$ \\
$\frac{3}{4}$ & $e^{-x}$ & \\
$1$ & $1/\Gamma(3/4) + \sqrt{\pi}(x/2)^{3/4}
\left[ {\bf L}_{1/4}(x)-I_{1/4}(x)\right]$
& $-1/(2 \Gamma(3/4)) x^{-2}$ \\ \hline
\end{tabular} \end{center}
\caption[Scaling function]{Scaling function $\Psi(a,x)$
at the Lifshitz point of the spherical model and their leading asymptotic
behaviour for $x\rar\infty$.
$I_n$ and $K_n$ are modified Bessel functions and ${\bf L}_n$ is a modified
Struve function. \label{TabFun}}
\end{table}

Comparison with Schr\"odinger invariance eq.~(\ref{ZweiPunkt}) shows agreement
for the case $D=6$, while the scaling function has a different form in the
other cases. Recall that the system given by the Hamiltonian $\cal H$
has a dispersion relation of the form
\BEQ
E^2 - \frac{1}{4m^2} k^4 = \left( E+\frac{1}{2m} k^2\right)
\left( E-\frac{1}{2m} k^2\right) =0
\EEQ
rather than $E= k^2/(2m)$ which was used for making the
Schr\"odinger-invariance predictions.
We can only expect to recover the results of Schr\"odinger invariance
if the propagator actually solves the free Schr\"odinger equation, and not just
a more general forth-order differential equation. We see this to be the case
for $D=6$ from table~\ref{TabFun} and concentrate on this from now on.

In order to test the prediction for two- and three-point correlations,
we consider several scaling fields as defined in table~\ref{TabSF}.
\begin{table}[htb]
\begin{center}
\begin{tabular}{|cc|c|c|c|} \hline
$\phi$ & & $x_{\phi}$ & ${\cal M}_{\phi}/{\cal M}_{\sig}$
& $\nu_{\phi}$ \\ \hline
$\sig_{\vec{a}}$ & $s_{\vec{a}}$ & $3/2$ & $1$ & $\cal A$ \\
$\eps_{\vec{a}}$ & $s_{\vec{a}} s_{\vec{a}\,'}$ & $3$ & $2$ &
$\sqrt{2}{\cal A}^2$\\
$\eta_{\vec{a}}$ & $s_{\vec{a}}^* s_{\vec{a}\,'}$ & $3$ & $0$ & ${\cal A}^2$\\
$\Sigma_{\vec{a}}$ & $s_{\vec{a}}s_{\vec{a}\,'}s_{\vec{a}\,''}$ & $9/2$ &
$3$ &  $\sqrt{6} {\cal A}^3$  \\ \hline
\end{tabular} \end{center}
\caption[Some scaling fields.]{Some scalar scaling fields arising at the
Lifshitz point of the spherical
model at $D=6$ and their scaling dimensions $x_{\phi}$, masses
${\cal M}_{\phi}$ in units of the mass ${\cal M}_{\sig}$ and the
normalization such that $\nu_{\phi}^{-1}\phi$ has $\Phi_0 =1$ in
eq.~(\ref{ZweiPunkt}). $\vec{a}\,'$ denotes
a space-time point on a neighboring site of the point given by $\vec{a}$.
\label{TabSF} }
\end{table}
We also give the scaling dimensions and the masses (in units of the mass of the
field $\sig$) of these fields. Concerning the field $\sig$, we can confirm
its values for $x_{\sig}$ and $\nu_{\sig}$ from eq.~(\ref{eq:DeuxP}).

The calculation of the other correlations is simple because, since
$\Lambda(-\vec{k}\,) = \Lambda(\vec{k}\,)$, the imaginary part of
$<s_a s_b^*>$ vanishes and we have
$C(\vec{a}-\vec{b})=<s_{\vec{a}}s_{\vec{b}}^*>$.

Consider the two-point functions first. (From now on, $a,b,c$ denote space-time
vectors and we also write $a=(\vec{r}_a,t_a)$ etc.)
Obviously, $<\eps_a>=0$. Then,
\BEA
<\eps_a \eps_b^*>&=&
<s_a s_{a'} s_b^* s_{b'}^*> = C(a-b) C(a'-b')+C(a-b')C(a'-b)
\nonumber \\
&\simeq& 2 [C(a-b)]^2 = 2 {\cal A}^4 t^{-2 x_{\sig}} e^{- r^2 /t}
\EEA
where in the second line the scaling limit was taken and we confirm the result
given in table~\ref{TabSF}. Next, we have $<\eta_a>=C(a-a')={\rm const}$. Then
the connected two-point function is
\BEQ
<\eta_a \eta_b>_c = <s_a^* s_{a'} s_b^* s_{b'}>_c = C(a-b') C(b-a') \simeq
t^{-2x_{\sig}} {\cal A}^4
\EEQ
whereas the exponential terms cancel. Finally, we have $<\Sigma_a>=0$ and
\BEQ
<\Sigma_{a} \Sigma_{b}^*> = < s_a s_{a'} s_{a''} s_b^* s_{b'}^* s_{b''}^*>
\simeq 6 [C(a-b)]^3 = 6{\cal A}^6 t^{-3 x_{\sig}} e^{- (3/2) r^2 /t}
\EEQ
and we have verified all entries in table~\ref{TabSF}. It is straightforward
to verify that
\BEQ
<\sig_a \eps_b^*> = <\sig_a \eta_b> = <\sig_a \Sigma_b^*> = <\eps_a \eta_b>
=<\eps_a \Sigma_b^*> = <\eta_a \Sigma_b^*> = 0
\EEQ
This fully confirms the Schr\"odinger two-point function eq.~(\ref{ZweiPunkt}).

At this stage, we clearly recognize the difference between the models described
by ${\cal H}_{SM}$ and $\cal H$, respectively, from the point of view of
Schr\"odinger (in fact already Galilean) invariance. The two fields $\eps$ and
$\eta$ have the same scaling dimension, but different masses. In the context
of the model described by ${\cal H}_{SM}$, however, these two distinct fields
get lumped together into just one field, $\widetilde{\eps}$, say. Correlations
of $\widetilde{\eps}$ will in general {\em not} satisfy Galilean invariance.
On the other hand, there is just a single order parameter field $\sig$.
Since in many applications one is only interested in the order parameter
correlations with itself, it is for that restricted purpose enough to stay
with the conventional form of ${\cal H}_{SM}$, rather than go to $\cal H$ with
the correct Galilean transformation properties.

We now turn to the three-point functions. To check the Bargmann superselection
rule, one may verify that, for example
\BEQ
<s_a s_b s_c^*> = <\eps_a \eps_b \eps_c^*> = <\eta_a \sig_b \eps_c^*>_c
= <\sig_a \sig_b \Sigma_c^*> = <\eps_a \eps_b \Sigma_c^*> = 0
\EEQ
A non-vanishing correlation is
\BEQ
<\sig_a \sig_b \eps_c^*> = <s_a s_b s_c^* s_{c'}^*> =
C(a-c) C(b-c') + C(a-c') C(b-c) \simeq 2 C(a-c) C(b-c)
\EEQ
Since $x_{\eps}=2 x_{\sig}$, this agrees indeed with the prediction
eq.~(\ref{DreiPunkt}) from Schr\"odinger invariance and we identify the
scaling function $\Psi_{\sig\sig,\eps}= \sqrt{2}$ which is a constant and
we have used the normalization given in table~\ref{TabSF}. Another check
of the Bargmann superselection rule is provided by showing that
$<\eps_a \sig_b \sig_c^*>=0$. Furthermore
\BEA
<\sig_a \eps_b \Sigma_c^*> &=& <s_a s_b s_{b'} s_c^* s_{c'}^* s_{c''}^*>
\simeq 6 C(a-c) [ C(b-c)]^2 \\
&=& 6 {\cal A}^6  (t_a - t_c)^{-x_{\sig}} (t_b -t_c)^{-2x_{\sig}}
e^{-(r_a-r_c)^2 /2(t_a-t_c)} e^{-(r_b-r_c)^2 /(t_b-t_c)} \nonumber
\EEA
which from table~\ref{TabSF} is seen to agree with (\ref{DreiPunkt}). For
normalized fields we identify $\Psi_{\sig\eps,\Sigma}=\sqrt{3}$. Apparently,
the massive scaling fields do reproduce the predictions of Schr\"odinger
invariance.

Finally, we look at some examples with the massless field
$\eta$ whose correlations
are not immediately zero. For example
\BEQ
<\eta_a \sig_b \sig_c^*>_c = <s_a^* s_{a'} s_b s_c^*>_c \simeq C(a-b) C(c-a)
\EEQ
We denote $\vec{a} = (\vec{r_a},t_a)$ and get
\BEQ
<\eta_a \sig_b \sig_c^*>_c = {\cal A}^4
(t_a -t_b)^{-x_{\sig}} (t_a -t_c)^{-x_{\sig}}
e^{-(r_a -r_b)^2 /2(t_a-t_b)} e^{+(r_a -r_c)^2 /2(t_a - t_c)}
\EEQ
which indeed agrees with (\ref{DreiPunkt}) (recall from eq.~(\ref{eq:LocTra})
that $\sig^*$ picks up a phase opposite to
$\sig$), since $x_{\eta}=2 x_{\sig}$.
Using normalized fields, we identify $\Psi_{\eta\sig,\sig}=1$. Next, consider
\BEA
<\eta_a \eps_b \eps_c^*>_c &=& <s_a^* s_{a'} s_b s_{b'} s_c^* s_{c'}^*>_c
\simeq 4 C(a-b) C(c-a) C(c-b) \label{eq:etetep} \\
&=& 4 {\cal A}^6 \left[ (t_a-t_b)(t_a-t_c)(t_b-t_c)\right]^{-x_{\sig}}
\exp\left[ - \frac{(r_a-r_b)^2}{2(t_a-t_b)}+\frac{(r_a-r_c)^2}{2(t_a-t_c)}
\right] \nonumber \\
& & \cdot e^{-(r_c-r_b)^2 /2(t_c-t_b)} \nonumber
\EEA
To relate this to the form (\ref{DreiPunkt}), we note that the first
line of the result (\ref{eq:etetep}) already takes the expected form and
we merely have to rewrite the last factor in the required scaling form.
Consider the argument of the scaling function $\Psi$
\BEQ
\frac{[(r_c-r_a)(t_b-t_a) - (r_b-r_a)(t_c-t_a)]^2}
{(t_b-t_a)(t_c-t_a)(t_c -t_b)}
= \frac{(r_c-r_b)^2}{t_c -t_b}+\frac{(r_c-r_a)^2}{t_c-t_a}
-\frac{(r_b-r_a)^2}{t_b-t_a}
\EEQ
The last two terms on the right-hand side can be rewritten as
\BEQ
\frac{(r_c-r_a)^2}{t_c-t_a}-\frac{(r_b-r_a)^2}{t_b-t_a}
= \frac{(r_c-r_a)^2}{t_c-t_a}
\left( 1 - \left(\frac{r_b-r_a}{r_c-r_a}\right)^2
\left( \frac{t_c-t_a}{t_b-t_a}\right)
\right) \rar 0
\EEQ
in the scaling limit. We thus find agreement with the expected form and
identify for the normalized fields $\Psi_{\eta\eps,\eps}(u)=
2 \exp(-u/2)$.
Finally, we consider
\BEA
<\eta_a \eta_b \eta_{c}>_c
&=& <s_{a}^* s_{a'} s_{b}^* s_{b'} s_{c}^* s_{c'}>_c \nonumber \\
&\simeq& C(a-b)C(c-a)C(b-c) + C(a-c)C(b-a)C(c-b) \\
&=& {\cal A}^6 \left[ (t_a-t_b)(t_c-t_a)(t_b-t_c)\right]^{-x_{\sig}}
\cdot \nonumber\\
&& \left\{ \exp\left( -\frac{1}{2} [ \Delta_{a,b}+\Delta_{c,a}+\Delta_{b,c}]
\right) + \exp\left( -\frac{1}{2} [ \Delta_{b,a}+\Delta_{a,c}+\Delta_{c,b}]
\right) \right\} \nonumber
\EEA
where $\Delta_{a,b}=(r_a-r_b)^2/(t_a-t_b)=-\Delta_{b,a}$. The very fact
that this correlation does not vanish again confirms that the
field $\eta$ is massless. We verify that
\BEQ
\Delta_{a,b}+\Delta_{c,a}+\Delta_{b,c} =
\frac{[ (r_a-r_c)(t_b-t_c)-(r_b-r_c)(t_a-t_c)]^2}{(t_a-t_b)(t_a-t_c)(t_b-t_c)}
\EEQ
and find, identifying $\Psi_{\eta\eta,\eta}(u)=2 \cosh(u/2)$, complete
agreement with the prediction eq.~(\ref{DreiPunkt}).

\subsection{Critical relaxation of the spherical model}

Having studied in some detail the case of a setting of infinite extent
in both time and space directions, we now look into one example
with a macroscopically prepared initial state which is not the
equilibrium state and the system is then allowed to relax towards
equilibrium. While for very long times, we are back to the dynamical
scaling considered so far, it was realized \cite{Jans89} that
already the intermediate stages of the relaxation process display
universal behaviour. The exponents and scaling functions describing
this have been calculated via an $\eps-$ expansion \cite{Jans89}.
Since we are merely interested in the special case of a dynamical
exponent $z=2$, we concentrate on the $n\rar\infty$ limit of the
$O(n)$ vector model. It was shown in \cite{Jans89} that for
a purely relaxational dynamics towards the equilibrium critical
point the exact response function is in $d\leq 4$ space dimensions
\BEQ
G_{\vec{k}}(t,t') = \Theta(t-t') \left( t/t'\right)^{1-d/4}
\exp\left[-\lmb k^2 (t-t')\right]
\EEQ
where $\Theta (t-t')$ is $1$ for $t>t'$ and zero otherwise.
Fourier transformation in space gives
\BEQ
G(\vec{r};t,t') = \Theta(t-t') (\pi/\lmb)^{d/2} t^{-d/2}
\left(\frac{t}{t'}\right)^{1-d/4} \left(1-\frac{t'}{t}\right)^{-d/2}
\exp\left[-\frac{1}{4\lmb^2}\frac{r^2}{t-t'}\right]
\EEQ
This is indeed consistent with the result eq.~(\ref{TOber}) for the
two-point function. We  identify $x=d/2$, ${\cal M}=1/(2\lmb^2)$ and
\BEQ
\Phi(v) = \Theta(v-1) (\pi\lmb)^{d/2} v^{1-d/4} \left(1 - v^{-1}\right)^{-d/2}
\EEQ
which also agrees with the consistency conditions given in section~3.

\section{Some remarks beyond $\theta=2$}

Since for generic values of the anisotropy exponent $\theta$ there is
at present no general approach available, we content ourselves with
a few results from selected models. In any case, these permit to
submit the conformal invariance result eq.~(\ref{eq:CRF}) \cite{Card85}
to a test.

\subsection{Lifshitz points of higher order in the spherical model}

We come back to the ANNNS model introduced earlier. Now, we add further
interaction terms along the axes denoted as ``space'' dimensions. Since we
are only interested here in the spin-spin correlation, it is sufficient
for us to consider the real Hamiltonian ${\cal H}_{SM}$ eq.~(\ref{eq:RSM}).
The Fourier transform of the couplings now is
\BEQ
J(\vec{k}\,) = 2J \left( \sum_{i=1}^{D} \cos k_i +
\sum_{j=1}^{d} \sum_{i=1}^{n} \kappa_i \cos[(i+1) k_j] \right)
\EEQ
Previously, we had taken $\kappa=\kappa_1$ to be the only non-vanishing
coupling. With several of the $\kappa_i$ non-zero, the phase diagram will
contain lines of Lifshitz points (also called Lifshitz points of
first order, where $L=2$) which end in a Lifshitz point of second
order (with $L=3$), in analogy to
the definition of multicritical points, see \cite{Selk92} for a review.
Lifshitz points of higher order are defined analogously. At a Lifshitz point of
order $L-1$, we have
\BEQ
J(\vec{k}\,) \simeq 2JD+d\sum_{i=1}^{n}
\kappa_i -\frac{1}{2} \sum_{j=d+1}^{D} k_{j}^2
- c_{L} \sum_{j=1}^{d} k_{j}^{2L} + \cdots
\EEQ
which defines the readily calculable constant $c_L$. When $d'=1$, the
anisotropy exponent is $\theta=L$. We are interested in the
critical correlation function $C(\vec{a}-\vec{b}\,)=$
$<\sig_a \sig_b>$. This can be calculated
exactly, see \cite{Frac93}. As we had already seen for Lifshitz points of
first order above, the correlations of the model considered here will
only in special cases solve the dispersion relation $E \sim k^{\theta}$,
rather than $E^2 \sim k^{2\theta}$. This should be the case if the scaling
function does not just show a power-law behaviour for large values of $u$.
This is the case if\footnote{Eq.~(2.22) in \cite{Frac93} contains a typing
error and correctly reads $a=\frac{1}{2}d -\frac{L-1}{2L} m -1 =
\frac{1}{2}(d-d_{-})$}
\BEQ
D = L+2+2m \;\; , \;\; m = 1,\ldots,L-1
\EEQ
Then the critical two-point function is in the scaling limit
$r\rar\infty, t\rar\infty$ with $u=r^{\theta}/t$ fixed \cite{Frac93}
\BEQ
C(\vec{r},t) = {\cal A}_{L,D} t^{-(2m+1)/L}\, \Xi\left( L,\frac{2m+1}{2L};
\frac{2^{1/L}}{4L c_{L}^{1/L}} \frac{r^2}{t^{2/L}} \right)
\EEQ
where $\Xi$ can be expressed as a finite sum of generalized hypergeometric
functions and ${\cal A}_{L,D}$ is a known non-universal constant.
Here we merely consider the behaviour for large values of the
scaling variable $u$, to leading order \cite{Frac93}
\BEA
C(\vec{r},t) &\simeq& {\cal B}_{L,D}\,
t^{-(2m+1)/L}\, u^{(D-3\theta)/(2\theta(\theta-1))} \nonumber \\
&\cdot& \exp\left[ {\cal C}_L
\cos\left(\frac{\pi}{2}\frac{L}{L-1}\right) u^{1/(\theta-1)} \right]
\label{eq:SMSkal} \\
&\cdot& \cos\left[ {\cal C}_L \sin\left(\frac{\pi}{2}\frac{L}{L-1}\right)
u^{1/(\theta-1)} + \frac{\pi}{2}\frac{L}{L-1}\frac{D-3L}{2L} \right]
\nonumber
\EEA
where ${\cal B}_{L,D}$ and ${\cal C}_L >0$ are known non-universal constants.
We remark that only the power prefactor of the scaling function depends
on $D$, the rest of the scaling function depends in the large $u$ limit only
on $L$.

Let us compare this result with the prediction eq.~(\ref{eq:CRF}) following
from conformal invariance \cite{Card85}. While conformal invariance gives
for a non-conserved order parameter a simple exponential behaviour for
the scaling function, we rather find
for the higher order Lifshitz points ($L\geq 3$) a stretched-exponential
behaviour. The two forms only agree for $\theta=2$.
On the other hand, for a conserved order parameter the
conformal invariance scaling function for $u$
large is of the form \cite{Card85}
$\psi(u) \sim u^{-2x/3z} e^{-u^{1/3}} \cos(\sqrt{3}u^{1/3})$ which for
$z=4$ is van Hove theory. It is interesting to note that for $L=4$, the
ANNNS model reproduces the same behaviour.

Finally, we look into the case where $\theta=1/2$. This is realized if
$d=1$ and $L=2$. Now the direction parallel to the next-to-nearest neighbors
interaction will be interpreted as ``time'' and the other direction
are referred to as ``space''. For $D=4$, we find again an exponential-like
behaviour for the spin-spin two-point function \cite{Frac93}
\BEQ \label{eq:halb}
C(\vec{r},t) \sim t^{-3/2}  u^{-3} \exp\left( -\frac{1}{2} u^{1/(\theta-1)}
\right)
\EEQ
in the scaling limit with $u=r^{1/2} /t$ fixed and $\theta=1/2$. Again, this
is different from the conformal invariance prediction eq.~(\ref{eq:CRF}).

\subsection{Directed percolation}

As a further example for a strongly anisotropic critical system
we consider directed percolation in $1+1$ dimensions (for a review,
see \cite{Kinz83}). In percolation, sites or bonds are filled at
random with probability $p$ and percolation proceeds along paths
between occupied nearest neighbor links. In directed percolation,
there is in addition a preferred direction and percolation is allowed
to proceed only in one sense along this direction. This preferred
direction is called ``time'' and the orthogonal ones ``space''.
Consider the pair connectedness $G(\vec{r},\vec{r}\,')$ which is a
measure of the probability that sites at $\vec{r}$ and $\vec{r}\,'$ are
connected by a percolating path. It is well known that there is
a critical value $p_c$ such that one has the scaling form
\BEQ \label{eq:Thalf}
G(r,t) = {\cal A} t^{-2\beta/\nu_{\|}} \Phi({\cal B} v)
\;\; , \;\; v = r/ t^{\nu_{\perp}/\nu_{\|}}
\EEQ
where $\beta,\nu_{\perp},\nu_{\|}$ are critical exponents, the
anisotropy exponent $\theta=\nu_{\|}/{\nu_{\perp}}$ and $r,t$ measure the
``space'' and ``time'' distances, respectively. Precise numerical values
on $p_c$ on various lattice  and of the exponent have been obtained,
see \cite{Benz84,Essa88}. For our purposes it is enough to notice that
in $(1+1)D$, to which we restrict ourselves here, we have
$\theta \simeq 1.5807$ \cite{Essa88}.

Benzoni \cite{Benz84} studied the pair connectedness by calculating numerically
the moments
\BEQ
\chi^{(n)} = {\cal A} \int_{-\infty}^{\infty} dv |v|^n \Phi( {\cal B} v)
\EEQ
and verified that the following ratios
\BEQ
C = \frac{[\chi^{(1)}]^2}{\chi^{(0)} \chi^{(2)}} \;\; , \;\;
F = \frac{[\chi^{(2)}]^2}{\chi^{(0)} \chi^{(4)}} \;\; , \;\;
G = \frac{[\chi^{(2)}]^2}{\chi^{(1)} \chi^{(3)}}
\EEQ
are independent of the non-universal scale factors ${\cal A},{\cal B}$
and should therefore be universal. A
careful numerical computation \cite{Benz84}
then yields numerical values for $C,F,G$ for various lattices of
both directed site and directed bond percolation. The results are in full
agreement with universality \cite{Benz84}.

We proceed to analyse these results in the following way. We try the
ansatz for the scaling function $\Phi(v)$
\BEQ \label{eq:BeAn}
\Phi(v) = |v|^b \exp(-v^a)
\EEQ
where $a,b$ are constants to be determined. Then
$\chi^{(n)}=2 a^{-1} \Gamma((n+b+1)/a)$. We now fit this form
to Benzoni's \cite{Benz84} numerical results for $C,F$ and $G$ and find
\BEA
a = 2.49 \pm 0.16 \;\; , \;\; b =-0.016 \pm 0.03 & &
\mbox{\rm from $C$} \nonumber \\
a = 2.58 \pm 0.12 \;\; , \;\; b =-0.023 \pm 0.03 & &
\mbox{\rm from $F$} \\
a = 2.62 \pm 0.17 \;\; , \;\; b =-0.023 \pm 0.07 & &
\mbox{\rm from $G$} \nonumber
\EEA
with the mean values $a=2.56(7)$ and $b=-0.02(3)$. Since the scaling
function $\Phi(v)$ is finite for $v=0$ \cite{Benz84}, we interpret this
result as implying that $b=0$. In table~\ref{TabBen}
\begin{table}[htb]
\begin{center}
\begin{tabular}{|c|ccccc|} \hline
ratio & 1 & 2 & 3 & 4 & 5 \\ \hline
$C$ & 2.41 & 2.56 & 2.49 & 2.24 & 2.39 \\
$F$ & 2.50 & 2.69 & 2.69 & 2.32 & 2.49 \\
$G$ & 2.54 & 2.74 & 2.65 & 2.37 & 2.54 \\ \hline
\end{tabular}
\end{center}
\caption[Exponent $a$]{Exponent $a$ as determined from the moment
ratios $C,F,G$ for the following realizations of directed percolation:
(1) square bond, (2) square site, (3) square site-bond, (4) triangular
bond, (5) triangular site. \label{TabBen} }
\end{table}
we give the
results for $a$ as found using the ansatz (\ref{eq:BeAn}) with $b=0$
from the ratios $C,F,G$ and various realizations of directed percolation.
Note that the estimates for $a$ obtained from
different moments and different
lattice realizations of directed percolation are the same, which means
that the chosen ansatz does indeed describe the available data.
{}From all this we conclude
\BEQ \label{N1}
a = 2.6 \pm 0.2
\EEQ
Making contact with our previous results, formulated in terms of the
scaling variable $u=v^{\theta}=r^{\theta}/t$, we obtain from (\ref{eq:BeAn})
\BEQ \label{N2}
a = \frac{\theta}{\theta-1} \simeq 2.72\ldots
\EEQ
using the known value of $\theta$. Comparison of (\ref{N1}) and (\ref{N2})
implies that also in this class of models
the two-point function scaling function appears to be consistent with
the {\em same} stretched-exponential form as already observed for
the Lifshitz points of the spherical model and in disagreement with
conformal invariance eq.~(\ref{eq:CRF}).

\section{Conclusions}

In this paper, we have examined the simplest consequences of the
hypothesis of local dynamical scaling with space-time dependent
local rescaling factors $\lmb(\vec{r},t)$. We have seen that for
the special case of an anisotropy (or dynamical) exponent $\theta=2$,
the Schr\"odinger group, which is the non-relativistic limit of the
conformal group, is a sensible candidate for a group
of local scale transformations.
The treatment of Schr\"odinger invariance (of quasiprimary fields)
is in many respects quite analogous to conformal invariance.
However, there are a few distinctions, the main one being the role of
the phase transformation which is not present in the conformal group.
We hope that the experience obtained in this
simplest non-conformal case may become useful for the extension of the
method to generic anisotropy exponents $\theta$. We have
derived the form of the two-point and three-point functions for both
infinite space and time (eqs.~(\ref{ZweiPunkt},\ref{DreiPunkt})) and
for the two-point function also if either space or time is restricted
to the half-infinite space (eqs.~(\ref{SOber},\ref{TOber})). The results
obtained are in agreement with and extend those following from the
weaker restrictions of Galilean invariance.

The Lie algebra of the Schr\"odinger group can be naturally extended to an
infinite-dimensional one. We have not solved the problem of how to use
this infinite algebra to calculate the critical exponents and the
scaling functions in the correlations which are left undermined in this
work. We hope to come back to this in the future.

Several exactly solvable statistical models with anisotropy exponent
$\theta=2$ were seen to reproduce the results of Schr\"odinger invariance
for the two- and three-point functions. In particular, we have seen
that due attention must be paid for correctly implementing the
changes of the phases of scaling fields as demanded by Galilean
invariance.

Evidence from some models with anisotropy exponent $\theta\neq 2$
suggests that, at least for large values of the scaling variable
$u=r^{\theta}/t$, the two-point scaling correlation function might
behave as
\BEQ
\Phi(u) \sim \exp\left( - u^{1/(\theta-1)}\right)
\EEQ
(where oscillating and power-law prefactors as well as non-universal
scale factors were suppressed). We have found examples for $\theta=n$,
with any integer $n \geq 3$, for $\theta\simeq 1.58\ldots$ and
for $\theta=1/2$ (for which the exact scaling function eq.~(\ref{eq:halb})
is known). This finding is in disagreement
with the form eq.~(\ref{eq:CRF}) suggested by using conformal
invariance in space.

\zeile{1.5}
\noindent{\bf Acknowledgements}
\zeile{1}
\noindent It is a pleasure to thank JL Cardy, M Droz, I Peschel, V
Rittenberg and G Sch\"utz for useful remarks and the D\'epartement de
Physique Th\'eorique of the Universit\'e de Gen\`eve, where this work was
done. This work was supported by the
SERC and
by the Swiss
National Science Foundation.

\newpage

\appendix                     %hier faengt der Anhang an
\appsection{A}{Solution of a system of linear differential equations}

We derive the general solution $H=H(r,s;\tau,\sig)$ of the following system of
differential equations
\BEA
\left( \tau \partial_{\tau} +\sig \partial_{\sig} +\frac{1}{2} r\partial_r
+\frac{1}{2} s\partial_s \right) H &=& 0 \nonumber \\
\left( \tau \partial_r + \sig \partial_s \right) H &=& 0  \label{eq:A1} \\
\left( \tau^2 \partial_{\tau} +\sig^2 \partial_{\sig} +\tau r\partial_r
+\sig s \partial_s \right) H &=& 0 \nonumber
\EEA
The technique consists of subsequent solution and resubstitution, see
\cite{Kamk59}. The second
of the eqs.~(\ref{eq:A1}) is solved by
\BEQ
H = \widetilde{H}(t; \tau,\sig) \;\; , \;\; t = \frac{r}{\tau}-\frac{s}{\sig}
\EEQ
while the other two equations become
\BEA
\left( \tau \partial_{\tau} +\sig\partial_{\sig}-\frac{1}{2} t \partial_t
\right) \widetilde{H} &=& 0 \nonumber \\
\left( \tau^2 \partial_{\tau} + \sig^2 \partial_{\sig} \right)
\widetilde{H} &=& 0
\EEA
The first of those is solved by
\BEQ
\widetilde{H} = \widetilde{\widetilde{H}}(u,v) \;\; , \;\;
u = \tau t^2 \;\; , \;\; v = \sig t^2
\EEQ
and the second one becomes
\BEQ
\left( u^2 \partial_u + v^2 \partial_v \right)
\widetilde{\widetilde{H}} = 0
\EEQ
with the solution $\widetilde{\widetilde{H}} =
\Psi( u^{-1} - v^{-1})$ where $\Psi$ is an arbitrary function.
Backsubstitution then yields the result eq.~(\ref{eq:HLoes}) in the text.

\newpage
\appsection{B}{Impossibility of non-conventional central extensions}

Consider the centrally extended (infinite) Schr\"odinger algebra
\BEA
\left[ X_n , X_m \right] &=& (n-m) X_{n+m} +\frac{c}{12}
\left( n^3 -n\right) \delta_{n+m,0}\nonumber \\
\left[ X_n , Y_m \right] &=& \left( \frac{n}{2} - m\right) Y_{n+m}
+ D(n,m) \nonumber \\
\left[ X_n , M_m \right] &=& -m M_{n+m} +E(n,m) \nonumber \\
\left[ Y_n , Y_m \right] &=& (n-m) M_{n+m} +A(n,m) \nonumber \\
\left[ Y_n , M_m \right] &=& F(n,m) \nonumber \\
\left[ M_n , M_m \right] &=& K n \delta_{n+m,0}
\EEA
where $A,D,E,F$ are numbers and $c,K$ are constants.
The special form of the central
extensions for $X_n, M_n$ is well known. We show that, with the only exception
of $c$, these central extensions
either have to vanish or can be reabsorbed into the generators.

This follows from the Jacobi identities. Begin with $D(n,k)$.
Consider $[X_n,[X_m,Y_k]]$ and
their cyclic permutations. This implies
\BEQ \label{Jd}
\left( \frac{m}{2}-k\right) D(n,m+k) - \left(\frac{n}{2}-k\right) D(m,n+k)
-(n-m) D(n+m,k) = 0
\EEQ
Let $n=0$. It follows (besides $D(0,0)=0$)
\BEQ \label{Jdd}
D(m,k) = -\frac{1}{2} \frac{m-2k}{m+k} D(0,m+k)
=: - \left( \frac{m}{2} - k\right) d(m+k)
\EEQ
which defines $d(k)$. Turning to $E(n,m)$, consider
$[X_n,[X_m,M_{\ell}]]$ and their cyclic permutations. This implies
\BEQ \label{eq:Je}
-\ell E(n,m+\ell) + \ell E(m,n+\ell) - (n-m) E(n-m,\ell) = 0
\EEQ
Now take $m=0$ in (\ref{eq:Je}) and then either
$\ell=0$ or $n+\ell=0$, implying
$E(n,0)=0$ for all $n$. We therefore write $E(n,m)=m{\cal E}(n,m)$.
Inserting in (\ref{eq:Je}) and taking $m=0$, we find that
${\cal E}(n,l)={\cal E}(0,n+\ell)+\delta_{n+\ell,0}\eps(n)
+\delta_{\ell,0}\widetilde{\eps}(n)$.
With the definitions ${\cal E}(0,n)=e(n)$ and $\eta(n)=-n\eps(n)$,
we have $E(n,m)=m e(n+m)+\eta(n)\delta_{n+m,0}$.
Backsubstitution into (\ref{eq:Je}) implies
\BEQ
(n+m)\left(\eta(n)-\eta(m)\right) = (n-m) \eta(n-m)
\EEQ
Let $n=0$ and get $\eta(n)=\eta(-n)$. Let $n+m=0$ to see
that $\eta(2n)=0$ for all $n$.
Taking $m=2n$, we get $\eta(n)=-\frac{1}{3}\eta(-n)$.
Consequently, $\eta(n)=0$ for all $n$.
Turning to $A(m,k)=-A(k,m)$, consider $[X_n,[Y_m,Y_k]]$ and
permutations. We get
\BEQ \label{Ja}
(m-k)E(n,m+k) +\left(\frac{n}{2}-k\right) A(n+k,m) -
\left(\frac{n}{2}-m\right) A(n+m,k) = 0
\EEQ
We use the result for $E(n,m)$, the antisymmetry of $A(m,k)$, insert in
(\ref{Ja}), put $n=0$ and divide by $m+k$ to get
$A(k,m) = (m-k) e(m+k)+a(k)\delta_{k+m,0}$, with $a(k)=-a(-k)$.
Backsubstitution into (\ref{Ja}) then implies $(3n/2 +m) a(-m) =
(n/2 -m) a(n+m)$. We now choose $n=2m$ and get $a(m)=0$.
To see that $K=0$, consider $[Y_n,[Y_m,M_k]]$ and its permutations to get
$(n-m) K n \delta_{n+m+k,0} = 0$.
Finally, we turn to $F(m,k)$ and consider
$[X_n,[Y_m,M_k]]$ and permutations to obtain
\BEQ \label{Jf}
k F(m,n+k) - \left(\frac{n}{2} -m\right) F(n+m,k) = 0
\EEQ
Put $n=0$ to find $F(m,k)=f(m) \delta_{m+k,0}$.
We now have to distinguish two cases. i) The index $m$ of $Y_m$ is
half-integer. Since the index $k$ of $M_k$ is always integer, we directly
have $F(m,k)=0$. ii) The index $m$ of $Y_m$ is integer. Backsubstitution
into (\ref{Jf}) then gives
\BEQ
(n+m)f(m) + \left( \frac{n}{2}-m \right) f(n+m) = 0
\EEQ
Let $n+m=0$ and find $f(0)=0$. Then, let $m=0$ and get $f(n)=-2f(0)=0$.

Consequently, the only surviving terms are given by $d(m)$ and $e(n)$. These
can be absorbed into the generators by
defining $\widetilde{M}_{n} = M_n - e(n)$ and $\widetilde{Y}_m = Y_m - d(m)$.
The only central term remaining is the one parametrized by $c$.
This proves the assertion.


\newpage

\begin{thebibliography}{99}

\bibitem{Bela84} A.A. Belavin, A.M. Polyakov and A.B. Zamolodchikov,
Nucl. Phys. {\bf B241}, 333 (1984)

\bibitem{Card90} J.L. Cardy, in C. Domb and J.L. Lebowitz (Eds)
{\em Phase Transitions and Critical Phenomena}, Vol. 11, Academic Press
(New York 1987), p. 55;\\
J.L. Cardy, in E. Br\'ezin and J. Zinn-Justin (Eds)
{\em Fields, Strings and Critical Phenomena}, Les Houches XLIX,
North Holland (Amsterdam 1990), p. 169; \\
C. Itzykson and J.M Drouffe, {\em Statistical Field Theory}, Vol. 2, ch. 9,
Cambridge University Press (Cambridge 1988); \\
P. Christe and M. Henkel, {\em Introduction to Conformal
Invariance and Its Applications to
Critical Phenomena}, Springer Lecture Notes in Physics,
New Series m: Monographs, Vol. m16, Springer (Berlin 1993)

\bibitem{Ferr67} R.A. Ferrell, N. Menyh\'ard, H. Schmidt, F. Schwabl and
P. Sz\'epfalusy, Phys. Rev. Lett. {\bf 18}, 891 (1967); \\
B.I. Halperin and P.C. Hohenberg, Phys. Rev. Lett. {\bf 19}, 700 (1967)

\bibitem{Hohe77} P.C. Hohenberg and B.I. Halperin, Rev. Mod. Phys.
{\bf 49}, 435 (1977)

\bibitem{Bind74} K. Binder and D. Stauffer, Phys. Rev. Lett.
{\bf 33}, 1006 (1974);\\
J. Marro, J.L. Lebowitz and M.H. Kalos, Phys. Rev. Lett.
{\bf 43}, 282 (1979);\\
B.D. Gaulin, S. Spooner and Y. Morii, Phys. Rev. Lett. {\bf 59}, 668 (1987)

\bibitem{Jans89} H.K. Janssen, B. Schaub and B. Schmittmann, Z. Phys.
{\bf B73}, 539 (1989)

\bibitem{Bray93} A.J. Bray, Physica {\bf A194}, 41 (1993)

\bibitem{Kinz83} W.Kinzel, in G. Deutscher, R. Zallen and J. Adler (Eds)
{\em Percolation Structures and Processes}, Adam Hilger (Bristol 1983), p. 425

\bibitem{Horn75} R.M. Hornreich, M. Luban and S. Shtrikman, Phys. Rev. Lett.
{\bf 35}, 1678 (1975)

\bibitem{Card85} J.L. Cardy, J. Phys. {\bf A18}, 2771 (1985)

\bibitem{Card93} J.L. Cardy, private communication

\bibitem{Ma76} S.-K. Ma, {\em Modern Theory of Critical Phenomena},
Benjamin (Reading MA, 1976)

\bibitem{Poly70} A.M. Polyakov, Sov. Phys. JETP Lett. {\bf 12}, 381 (1970)

\bibitem{Droz91} M. Droz, L. Frachebourg and M.C. Marques, J. Phys.
{\bf A24}, 2869 (1991)

\bibitem{Nied72} U. Niederer, Helv. Phys. Acta {\bf 45}, 802 (1972)

\bibitem{Hage72} C.R. Hagen, Phys. Rev. {\bf D5}, 377 (1972)

\bibitem{Baru73} A.O. Barut, Helv. Phys. Acta {\bf 46}, 496 (1973)

\bibitem{Perr77} M. Perroud, Helv. Phys. Acta {\bf 50}, 233 (1977)

\bibitem{Nied74} U. Niederer, Helv. Phys. Acta {\bf 47}, 167 (1974)

\bibitem{Nied78} C.-C. Wei, Y.-C. Yang, H.J. Annegarn, R.J. Yeh and
C.-Y. Wang, in H.-D. Doebner, J.-D. Henning and T.D. Pavlev (Eds)
{\em Group Theoretical Methods in
Physics}, Springer Lecture Notes in Physics, vol. 313,
Springer (Berlin 1988), p. 282;\\
U. Niederer, Helv. Phys. Acta {\bf 51}, 220 (1978)

\bibitem{Beck91} J. Beckers, N. Debergh and A.G. Nikitin, J. Phys. {\bf A24},
L1269 (1991)

\bibitem{Jack80} R. Jackiw, Ann. of Phys. {\bf 129}, 183 (1980);
{\bf 201}, 83 (1990)

\bibitem{Duva91} C. Duval, G. Gibbons and P. Horvathy, Phys. Rev.
{\bf D43}, 3907 (1991)

\bibitem{Henk92} M. Henkel, Int. J. Mod. Phys. {\bf C3}, 1011 (1992)

\bibitem{Levy67} J.-M. Levy-Leblond, Comm. Math. Phys. {\bf 4}, 157 (1967);
{\bf 6}, 286 (1967)

\bibitem{Barg54} V. Bargmann, Ann. of Math. {\bf 59}, 1 (1954)

\bibitem{Kreu81} H.J. Kreuzer, {\em Nonequilibrium Thermodynamics and
its Statistical Foundations}, Clarendon Press (Oxford 1981)

\bibitem{Osbo93} H. Osborn and A. Petkos, Cambridge preprint DAMTP/93-31

\bibitem{Card84} J.L. Cardy, Nucl. Phys. {\bf B240}[FS 12], 514 (1984)

\bibitem{Kamk59} E. Kamke, {\em Differentialgleichungen: L\"osungsmethoden und
L\"osungen}, Vol. 2, 4$^{th}$ edition, Akademische Verlagsgesellschaft
(Leipzig 1959)

\bibitem{Ruja87} P. Ruj\`an, J. Stat. Phys. {\bf 49}, 139 (1987); \\
A. Georges and P. Le Doussal, J. Stat. Phys. {\bf 54}, 1011 (1989)

\bibitem{Kand90} D. Kandel, E. Domany and B. Nienhuis, J. Phys.
{\bf A23}, L755 (1990)

\bibitem{Berg92} O. Bergman, Phys. Rev. {\bf D46}, 5474 (1992)

\bibitem{Glau63} R.J. Glauber, J. Math. Phys. {\bf 4}, 294 (1963)

\bibitem{Gunt83} J.D. Gunton and M. Droz, {\em Introduction to the
Theory of Metastable and Unstable States}, Springer Lecture Notes in Physics,
vol. 183, Springer (Berlin 1983)

\bibitem{Sing83} S.S. Singh and R.K. Pathria, Phys. Rev. {\bf B31}, 4483 (1983)

\bibitem{Schu93} G. Sch\"utz and S. Sandow, Weizmann preprint
WIS-93/46/May.-PH,
Phys. Rev. E in press

\bibitem{Alca93} F.C. Alcaraz, M. Droz, M. Henkel and V. Rittenberg,
Geneva preprint UGVA-DPT 1992/12-799, to appear in Ann. of Phys.

\bibitem{Sigg77} E. Siggia, Phys. Rev. {\bf B16}, 2319 (1977)

\bibitem{Pokr80} V.L. Pokrovsky and A.L. Talapov, Sov. Phys. JETP {\bf 51},
134 (1980)

\bibitem{Selk92}  W. Selke, in C. Domb and J.L. Lebowitz (Eds)
{\em Phase Transitions and Critical Phenomena}, Vol. 15, Academic
(New York 1992), p. 1

\bibitem{Joyc72} G.S. Joyce, in C. Domb and M.S. Green {\em Phase Transitions
and Critical Phenomena}, Vol. 2, Academic (New York 1972), p. 375

\bibitem{Frac93} L. Frachebourg and M. Henkel, Physica {\bf A195}, 577 (1993)

\bibitem{Benz84} J. Benzoni, J. Phys. {\bf A17}, 2651 (1984)

\bibitem{Essa88} J.W. Essam, A.J. Guttmann and K. De'Bell, J. Phys. {\bf A21},
3815 (1988)

\end{thebibliography}
\end{document}